\documentclass[lettersize,journal]{IEEEtran}
\usepackage{amsmath,amsfonts}
\usepackage{algorithmic}
\usepackage{algorithm}
\usepackage{array}
\usepackage[caption=false,font=normalsize,labelfont=sf,textfont=sf]{subfig}
\usepackage{textcomp}
\usepackage{stfloats}
\usepackage{url}
\usepackage{verbatim}
\usepackage{graphicx}
\usepackage{cite}
\usepackage{float}
\usepackage{booktabs}
\usepackage{siunitx}
\usepackage{colortbl}
\usepackage{xcolor}
\usepackage{soul}
\usepackage{hyperref}
\definecolor{effort}{HTML}{6DFCFF}
\definecolor{outcome}{HTML}{F4CCCC}
\definecolor{person}{HTML}{C99AF7}
\usepackage{orcidlink}

\begin{document}

\title{Improving Automated Feedback Systems for Tutor Training in Low-Resource Scenarios through Data Augmentation}

\author{Chentianye Xu\orcidlink{0009-0007-3444-1719}, Jionghao Lin\orcidlink{0000-0003-3320-3907}, Tongshuang Wu\orcidlink{0000-0003-1630-0588}, Vincent Aleven\orcidlink{0000-0002-1581-6657}, and Kenneth R. Koedinger\orcidlink{0000-0002-5850-4768}
        % <-this % stops a space
\thanks{All the authors are with School of Computer Science, Carnegie Mellon University, Pittsburgh, PA, 15213 USA. Jionghao Lin is also with The University of Hong Kong. Corresponding author: Jionghao Lin (jionghao@hku.hk)}% <-this % stops a space
\thanks{This paper is an extended version of the conference paper Jionghao Lin, Eason Chen, Zifei Han, Ashish Gurung, Danielle R. Thomas, Wei Tan, Ngoc Dang Nguyen, \& Kenneth R. Koedinger. (2024). How Can I Improve? Using GPT to Highlight the Desired and Undesired Parts of Open-ended Responses. Proceedings of the 17th International Conference on Educational Data Mining, 236--250. https://doi.org/10.5281/zenodo.12729804}
}


\maketitle

\begin{abstract}
Tutoring is an effective instructional method for enhancing student learning, yet its success relies on the skill and experience of the tutors. This reliance presents challenges for the widespread implementation of tutoring, particularly in training novice tutors. To support tutor training programs, real-time automated feedback systems are essential for efficiently training large numbers of tutors. Lin et al.'s previous study employed Generative Pre-Trained Transformers (GPT) for sequence labeling to identify desirable and undesirable praise components in a tutor training dataset, providing explanatory feedback.
 However, this approach requires a significant amount of labeled data for fine-tuning, which is both labor-intensive and dependent on expert input. To address the challenges associated with extensive data labeling, the current study explores the use of prompting more advanced GPT models like GPT-4o to generate synthetic datasets for augmenting labeled response data, followed by fine-tuning a GPT-3.5 model. Our results demonstrate that: (1) fine-tuning with augmented dataset with size 520 generated by GPT-4o significantly improves GPT-3.5's performance in identifying praise components, with $F_{2}$ increasing by 20.5\% for effort-based praise and 17.3\% for outcome-based praise compared to the same model fine-tuned without augmentation. These improvements are also reflected in traditional metrics like M-IoUand IoU score.
(2) Our data augmentation approach generalizes effectively to identify other types of praise (i.e., person-based praise), with $F_{2}$ Score increasing by 16.5\% for person-based praise, alongside a 19.9\% increase in M-IoU and a 21.6\% increase in IoU with augmented dataset size 520, compared to the same model fine-tuned without augmentation. These findings suggest that for data-intensive tasks, synthetic data generated through GPT model prompting can substantially enhance fine-tuned model performance in low-resource scenarios. This approach offers a scalable and cost-effective solution to alleviate the reliance on extensive expert-labeled data, enabling broader adoption of automated tutor training systems and facilitating the rapid development of skilled tutors at scale.
\end{abstract}

\begin{IEEEkeywords}
Tutor training, large language models, data augmentation, sequence labeling, automated feedback system.
\end{IEEEkeywords}

\section{Introduction}
\IEEEPARstart{H}{uman} tutoring is an effective instructional method known for its efficacy in enhancing student learning. However, several logistical challenges hinder its widespread implementation, including the recruitment, training, and retention of skilled tutors~\cite{thomas2023tutor}. Training tutors is resource-intensive and often requires hands-on mentorship. Equipping novice tutors with effective tutoring strategies is a key aspect of this training. For example, rather than merely correcting an incorrect answer, skilled tutors engage students to uncover underlying misconceptions, enabling more effective support. Traditionally, these insights are imparted through direct training from experienced tutors, but this approach is limited by the shortage of expert tutors and the high cost of personalized mentorship ~\cite{lin2023using, hirunyasiri2023comparative, kakarla2024using, lin2024can}.

In response to the increasing demand for scalable hands-on support in tutor training, researchers are increasingly leveraging automated feedback systems \cite{borchers2024combining, wang2023chatgpt, lin2024can, lin2023learner, beck2008does, gurung2023common, patikorn2020effectiveness}. 
Many AI algorithms generate automated feedback \cite{cavalcanti2021automatic}, but their specific use in tutor training remains under-explored. AI could improve scalability and effectiveness, addressing the limitations of resource-intensive hands-on training. Traditional models like BERT \cite{devlin2019bert} face challenges due to limited labeled training datasets and  limited generalization ability, which constrain their ability to provide precise feedback \cite{lin2023using, ezen2020comparison}. Recent advances in Large Language Models (LLMs) offer a promising solution (e.g.,\cite{lin2024can}) to challenges such as limited labeled training datasets and poor model robustness by leveraging their vast knowledge base and strong generalization capabilities \cite{yang2024unveiling, zheng2024large}. These LLMs can dynamically adapt to domain-specific scenarios like generating educational feedback, making them well-suited for developing real-time feedback systems for tutor training \cite{lin2024can, yang2024unveiling}.

Fine-tuning LLMs yields accurate educational feedback but demands large amounts of labeled data, which is labor-intensive and requires specialized expertise \cite{lin2024can}. This challenge is especially noticeable in situations with limited data (low-resource scenarios), which makes the fine-tuning process more difficult. A promising method to address the challenge of low-resource data is text data augmentation using natural language processing techniques, which can generate synthetic data samples and expand the training dataset. 

While previous research \cite{cochran2023improving} has explored using LLMs like GPT-3.5 for text data augmentation in educational contexts, our study builds on these efforts by employing more advanced models such as GPT-4o to generate synthetic datasets specifically tailored for tutor training scenarios. Our data augmentation approach not only introduces more diverse linguistic variations but also maintains key aspects such as grammaticality and coherence, making the synthetic data particularly well-suited for generating high-quality, contextually relevant data. By leveraging LLMs, we aim to overcome the limitations \cite{feng2021survey} of traditional augmentation methods (e.g., random word swapping and synonym replacement) and generate richer datasets that improve model performance. Furthermore, ensuring that LLM-based data augmentation effectively generalizes is critical for successfully generating diverse types of educational feedback. Generalizing LLM-based data augmentation is not straightforward due to the diverse nature of educational responses, which can vary significantly in context, intent, and linguistic complexity. To this end, our study also aims to examine the generalizability of LLM-based data augmentation across multiple categories in the automated feedback generation process. In light of these considerations, we propose the following two \textbf{R}esearch
\textbf{Q}uestions (\textbf{RQ}s):

\vspace{3mm}

\noindent \textbf{RQ1:}  How effectively can data augmentation approaches improve the performance of fine-tuned LLMs in delivering explanatory feedback within low-resource scenarios?

\vspace{2mm}

\noindent \textbf{RQ2:}  To what extent can our proposed data augmentation approach generalize to other types of educational feedback?
\vspace{3mm}

 To address \textbf{RQ1}, we employed ChatGPT-4o and ChatGPT-3.5 to perform data augmentation on a limited set of labeled data and used the augmented data to fine-tune ChatGPT-3.5 and ChatGPT-4o, focusing on effort-based praise and outcome-based praise. The results showed that increasing the augmented training set size led to improved model performance across all metrics for both effort-based and outcome-based praise.
For \textbf{RQ2}, we extended our method to person-based praise to evaluate its generalizability. The results demonstrated that out data augmentation method generalize well to person-based praise.

\section{Background}
\subsection{Tutoring Practice: Giving Effective Praise}
Effective tutoring is crucial for improving student learning by combining academic knowledge with the ability to meet students' socio-motivational needs \cite{dietrichson2017academic, guryan2023not, nickow2020impressive, lin2022good}. However, training tutors to develop tutoring skills is challenging because they often lack hands-on learning opportunities that allow them to practice real-life scenarios as part of their professional development \cite{chine2022development}. The lack of practical experience hinders tutors from applying effective tutoring practices that support student learning. Consequently, it is necessary to develop tutor training programs that provide practical training, including the social-emotional and motivational dimensions of the tutoring process  \cite{chine2022development, reich2022teaching}. Our study focuses on the effective delivery of praise, a key component of human tutoring that significantly boosts student motivation, engagement, and learning outcomes \cite{jenkins2015rates, kamins1999person, thomas2023tutor}. %Research suggests that effective praise should be sincere, specific, immediate, and avoid repetitive phrases like \textit{``great job''}, and focus on the learning process such as  ``\textit{hard work paid off}''\cite{thomas2023tutor}.

The literature \cite{kamins1999person, thomas2023tutor, chine2022scenario, chine2022development} identifies three primary types of praise: effort-based, outcome-based, and person-based. Effort-based praise, which emphasizes the student's learning process (e.g., \textit{``I love the effort you put into this writing...''}), is considered the most desirable form of praise. Outcome-based praise acknowledges specific achievements (e.g., high scores or correct problem-solving) and often includes generic phrases such as \textit{``Good job!''} but is viewed as less desirable. Person-based praise, which attributes success to inherent qualities (e.g., saying \textit{``You are smart!''}), is the least desirable because it focuses on fixed traits \cite{kamins1999person}. To enhance student learning outcomes, it is essential to train tutors to consistently provide effective forms of praise, particularly effort-based praise. Prior research indicates that providing this type of praise can have a significant positive impact on student motivation and engagement \cite{kamins1999person, thomas2023tutor, chine2022scenario, chine2022development}. However, tutors may not always recognize when their responses contain ineffective praise, which highlights the importance of offering explanatory feedback to improve their skills \cite{chine2022development, thomas2023tutor, lin2024can, lin2024rephrase}.

Manually crafting feedback for tutor trainees, especially in identifying the types of praise used, is a labor-intensive process that requires substantial time and resources from expert tutors \cite{cavalcanti2021automatic}. This underscores the need for scalable, automated feedback systems that can assist tutor training programs by efficiently identifying and classifying tutor responses to provide explanatory feedback. By developing a LLM-based system to automate this process, we aim to reduce the workload involved in feedback generation and ensure timely, scalable support for tutors in training. Such systems would enhance the ability of tutors to consistently use effective praise strategies.

\subsection{Automated Feedback for Tutor Training}
Feedback is widely acknowledged for its profound impact on learning outcomes \cite{patikorn2020effectiveness, gurung2023common, gurung2023identification, lin2023learner, henderson2019impact}, with its effectiveness varying based on the content and delivery method. Hattie and Timperley \cite{hattie2007power} emphasize that the effectiveness of feedback depends on its relevance to the learning context, its timing, and its focus on addressing misconceptions or errors in reasoning. Immediate, explanatory feedback, which explains the reasons behind correct or incorrect responses, is particularly crucial for fostering active engagement and reflective practice among learners \cite{ryan2021designing, lin2023learner, hattie2007power, henderson2019impact}. This recognition of feedback's importance has led to the increased use of automated feedback systems in educational environments, such as OnTask, which enables educators to provide scalable feedback based on students' academic activities and performance through conditional rules \cite{pardo2018ontask}. However, the use of such systems in tutor training has not been extensively explored. One effective way to implement automated feedback in tutor training is through templated feedback \cite{lin2023using, lin2024can, lin2024rephrase}. Templated feedback, which includes specific references to desirable and undesirable elements of tutor responses, is informed by previous research demonstrating the benefits of a data-driven error diagnosis taxonomy for template-based feedback \cite{aleven2001towards}. Our study aims to utilize natural language processing (NLP) techniques to automate the identification of these key elements within tutor responses thereby facilitating the provision of templated explanatory feedback.

\subsection{Sequence Labeling for Feedback Generation}
Sequence labeling is a crucial task in NLP, essential for identifying and categorizing key text segments according to predefined labels \cite{jurafsky2024speech}. One illustrative subtask of sequence labeling is Named Entity Recognition (NER), which closely aligns with our study's objectives. NER automatically detects and classifies named entities—words or phrases with specific attributes—into categories such as person, organization, and location \cite{jurafsky2024speech, li2020survey}. For instance, in the sentence \textit{``Sarah mentioned that London becomes even more mysterious in the fall.''}, the terms \textit{``Sarah''}, \textit{``London''}, and \textit{``fall''} would be labeled as \texttt{Person}, \texttt{Location}, and \texttt{Time}, respectively, demonstrating NER's capability to distinguish and categorize entities within text.

Our study extends the application of sequence labeling to identify and highlight components of praise from tutor responses. This involves detecting specific words or phrases that indicate the kind of praise being used, thus providing tutors with insights into their feedback practices. For instance, in the phrase \textit{``You did a \underline{great job}''}, the term \textit{``great job''} is identified as outcome-based praise. By leveraging sequence labeling, we aim to develop an AI model that can highlight key components of praise, enabling the provision of automated explanatory feedback. An example of such feedback is, \textit{``Saying `great job' praises the student for the outcome. Consider focusing on the student's effort and process towards learning. Would you like to try responding again?''}. However, previous studies have identified two key challenges: (1) traditional models, such as BERT \cite{lin2023using}, was constrained by limited access to extensive datasets, hindering its ability to accurately identifying and categorizing feedback elements, and (2) The labor-intensive nature of sequence annotation, which underscores the necessity of developing more context-aware data augmentation methods. These challenges highlight the need for advanced NLP techniques to facilitate the provision of precise and informative feedback to tutors, particularly in low-resource scenarios.
% While previous research has employed sequence labeling techniques for similar purposes \cite{lin2023using}, the accuracy of these models in precisely identifying and categorizing feedback elements remains a challenge. This highlights the need for utilizing more advanced models to deliver accurate, informative feedback to tutors, thus enhancing their ability to provide effective praise.

% \subsection{Fine-tune Large Language Models in Education}
Recent advancements in NLP have underscored the potential of LLMs, such as GPT-3.5 and GPT-4 \cite{brown2020language, achiam2023gpt}, in various educational contexts through techniques like prompting and fine-tuning \cite{kasneci2023chatgpt}.  These models have demonstrated significant promise in enhancing the identification and categorization of key text segments \cite{devlin2018bert, brown2020language, achiam2023gpt}, which is critical for developing automated systems that provide targeted, explanatory feedback. Our previous work \cite{lin2024can} investigated the use of prompting and fine-tuning GPT models to identify both desired and undesired components in tutor open-ended responses involving praise. This study \cite{lin2024can} evaluated the models' capabilities in developing an automated system for providing explanatory feedback to tutors, and found that fine-tuning significantly outperformed prompting. Fine-tuning GPT models has shown considerable potential in educational applications \cite{kasneci2023chatgpt}. Fine-tuning involves adjusting the model's neural network to better fit specific domains, thereby enhancing its performance in those contexts \cite{jurafsky2024speech}. Building on our prior work, we aim to further enhance model performance in low-resource scenarios by leveraging data augmentation techniques, extending our previous findings and methods.

 % For instance, Latif and Zhai \cite{latif2024fine} employed a fine-tuned GPT-3.5 model for automatic scoring in science education, finding that it outperformed the established BERT model \cite{devlin2018bert} and demonstrated superior accuracy in various science education tasks. This illustrates the value of fine-tuning GPT models for educational purposes, providing precise and scalable solutions across diverse educational settings. Similarly, Bhat \textit{et al.} \cite{bhat2022towards} developed a method for generating assessment questions from text-based learning materials using a fine-tuned GPT-3 model. Human experts evaluated these generated questions for their educational usefulness, and the results were favorably received.

% Inspired by these studies, our research aims to apply fine-tuning methods to GPT models for generating explanatory feedback. While previous studies \cite{devlin2018bert, latif2024fine} have not directly addressed explanatory feedback, their success in using fine-tuned LLMs within educational domains indicates a promising direction for our work. By tailoring GPT models to the specific requirements of educational feedback, we anticipate significant advancements in automating and improving the feedback process. This effort will contribute to the growing body of evidence supporting the integration of fine-tuned LLMs in educational technology, potentially transforming the way feedback is generated and utilized in learning environments.

\subsection{Text Data Augmentation}
Data augmentation in NLP has gained significant interest, particularly in low-resource domains and new task. Data augmentation in NLP refers to techniques used to artificially expand a dataset by creating new examples through transformations of the original text such as synonym replacement, back-translation, or noise injection \cite{shorten2021text}. These methods help improve model performance, particularly in low-resource settings, by increasing the diversity of training data \cite{shorten2021text}. Despite its growing importance, data augmentation in NLP is still relatively underexplored because language data is composed of discrete tokens (words or characters), making it challenging to apply transformations without altering the meaning. For example, techniques like synonym replacement or paraphrasing can inadvertently change the context or nuance of a sentence. Unlike continuous data like images, where simple transformations (e.g., rotation or scaling) often preserve the content, language data requires more careful manipulation to maintain its original meaning and relevance. Therefore, NLP data augmentation methods often involve more sophisticated strategies to maintain the syntactic and semantic integrity of the augmented text. Some traditional data augmentation techniques for NLP include rule-based methods like synonym replacement, random insertion, deletion, and swapping of words, which have improved performance on text classification tasks, such as sentiment analysis and subjectivity detection. For example, Wei and Zou \cite{wei2019eda} used simple token-level perturbations to enhance sentiment analysis and product review classification. Sennrich et al. \cite{sennrich2016edinburgh} applied backtranslation to improve neural machine translation. Kobayashi \cite{kobayashi2018contextual} used contextual augmentation with pre-trained language models, such as BERT and GPT-2, to generate more contextually consistent text. Despite the effectiveness of these methods, traditional augmentation approaches can lead to semantic shifts and inconsistencies in the generated text. For instance, a synonym replacement might change the meaning of a sentence in subtle ways, such as altering the nuance or context. In contrast, leveraging LLMs for data augmentation allows for the generation of more semantically coherent and contextually consistent data, making them a better solution  than traditional methods \cite{long2024llms}. For instance, Ghosh et al. \cite{ghosh2023aclm} employ selective denoising to generate coherent augmentations for complex NER, addressing context-entity mismatches that traditional methods struggle with. Xiao et al. \cite{xiao2023freeal} explore human-free active machine learning method using LLMs as annotators to reduce annotation costs while maintaining high-quality data generation. Li et al. \cite{li2024empowering} propose Self-LLMDA framework, which automates instruction generation and selection for task-specific data augmentation. This approach \cite{li2024empowering} significantly enhances data quality by balancing generative breadth with task-specific precision.

Recent advancements in LLMs have also explored the approach of using powerful LLMs to generate task-specific training data, which subsequently improves the performance of less advanced LLMs \cite{khattab2023dspy, viswanathan2023prompt2model}. This approach is related to the concept of knowledge distillation\textemdash a process where a larger, more powerful model (often referred to as the ``teacher model'') is used to transfer knowledge to a smaller, more efficient model (referred to as the ``student model'') \cite{hinton2015distilling}. By using a high-capacity teacher model to create high-quality synthetic datasets, knowledge distillation aims to improve the capabilities of the student model, even when computational resources are limited.  This process allows the student model to approximate or replicate the performance of the teacher while being computationally more efficient. In this way, even though the student model has fewer parameters or less capacity than the teacher, the student model can still achieve high performance by leveraging insights gained from the teacher model's outputs. It is worth noting that the knowledge distillation approach has proven particularly effective compared to traditional data augmentation methods, as it allows for the creation of more fluent text \cite{viswanathan2023prompt2model}. Unlike traditional augmentation, which often relies on simple transformations and may inadvertently change the meaning of text, the use of LLMs allows for generating task-specific examples that are tailored to the particular knowledge components and linguistic features necessary for the student model to learn effectively \cite{long2024llms}.

\section{Method}
\subsection{Dataset}
\label{dataset}
Our study received ethical approval from the Institutional Review Board (IRB) at Carnegie Mellon University. For \textbf{RQ1}, we used the same dataset as our previous study \cite{lin2024can} to maintain consistency and facilitate comparison. This dataset includes responses from 65 volunteer tutors who participated in the \textit{Giving Effective Praise} lesson. The demographic breakdown of these tutors was: 52\% identified as White, 18\% as Asian, 52\% were male, and over half were aged 50 years or older. The purpose of the \textit{Giving Effective Praise} lesson is to instruct tutors on the effective use of praise to boost student motivation. We collected a total of 129 responses from tutors who completed the lesson, categorizing each response by the type of praise used (i.e., effort-based praise and outcome-based praise).  In our previous work, we partitioned a dataset of 129 labeled responses into a training set of 64 responses (44 containing effort-based praise and 23 containing outcome-based praise) and a test set of 65 responses (47 containing effort-based praise and 33 containing outcome-based praise). In this study, we examine low-resource settings by utilizing only 10\% of the entire dataset—namely, 13 labeled responses. Notably, all 13 responses in this reduced training set include both effort-based and outcome-based praise. The dataset details are also shown in Table \ref{tab:dataset_summary}. This extreme reduction is intended to simulate real-world situations where the availability of labeled data is severely limited. We then apply data augmentation techniques to this reduced subset. The choice of 10\% was made to ensure that there is sufficient initial information available to guide the augmentation process. Using a smaller subset would severely limit the semantic richness of the initial data, making it difficult for the augmentation methods to generate diverse and meaningful examples. We aim to strike a balance between simulating a low-resource scenario and providing enough foundational data to enable the augmentation techniques to produce useful, varied, and contextually appropriate examples.

\begin{table}[H]
\centering
\caption{Summary Statistics of Datasets for RQ 1}
\resizebox{\linewidth}{!}{
\begin{tabular}{llccc}
\toprule
\textbf{Research Question} & \textbf{Dataset Split} & \textbf{Total Responses} & \textbf{Effort-based} & \textbf{Outcome-based} \\
\midrule
{RQ1} & Training Set (13 subset) & 13 & 13 & 13 \\
                     & Test Set                 & 65 & 47 & 33 \\
\midrule
{RQ1 full} & Training Set            & 64 & 44 & 23  \\
                         & Test Set                & 65 & 47 & 33  \\
\bottomrule
\end{tabular}
}
\label{tab:dataset_summary}
\end{table}

For \textbf{RQ2}, we aim to evaluate the generalization ability of our data augmentation method. To do this, we collected an additional 10 labeled responses that exclusively feature person-based praise. These new responses are combined with our existing dataset of 129 labeled responses, which do not include person-based praise (the only response containing person-based praise had that part removed, so it now contains only effort-based praise), resulting in a total of 139 labeled responses. Given the data imbalance (only 10 out of 139 responses feature person-based praise), we treat this as a two-class NER problem. Instead of distinguishing between effort-based praise and outcome-based praise as we did in \textbf{RQ1}, we simplify the task to identify person-based praise versus all other content. This approach allows us to specifically test the model's ability to generalize our data augmentation technique to a new type of praise, thereby assessing its robustness and applicability to varied feedback scenarios. Similar to our approach in \textbf{RQ1}, we split the 139 responses into a training set and a test set. The training set comprises 70 labeled responses, of which 5 include person-based praise. The test set consists of 69 labeled responses, also with 5 featuring person-based praise. For the fine-tuning stage, we used all 70 real data points in the training set, including both person-based praise and non-person-based praise responses. This approach differs from RQ1, where we used 10\% (13 instances) of the whole dataset, because person-based praise is inherently a low-resource category. With only 5 examples of person-based praise in the training set, using the full set was necessary to ensure the model could adequately recognize this category. The remaining 65 non-person-based praise responses were included to enhance the model's robustness across diverse contexts.

\subsection{Sequence Labeling}
Our goal is to provide explanatory feedback that highlights the components of effort-based, outcome-based, and person-based praise within tutor responses. To achieve this, we employed a sequence labeling approach. This sequence labeling approach is crucial for delivering automated explanatory feedback, as this approach allows for precise identification of praise types and the specific words of praise. By highlighting different types of praise from tutor trainees' responses, they can better understand their responses to open-ended questions in training, enabling them to refine their approach effectively.

Drawing from studies \cite{thomas2023tutor, chine2022scenario, chine2022development}, we developed annotation guidelines and specific examples of effort-based, outcome-based and person-based praise. Two expert educators, who first completed the \textit{Giving Effective Praise} lesson on our platform, were then tasked with annotating 129 tutor responses to identify attributes of effort-based and outcome-based praise. The current study followed the same annotation process and annotated 10 additional responses featuring person-based praise.

\begin{figure}[h]
\centering
\includegraphics[width=0.2\textwidth]{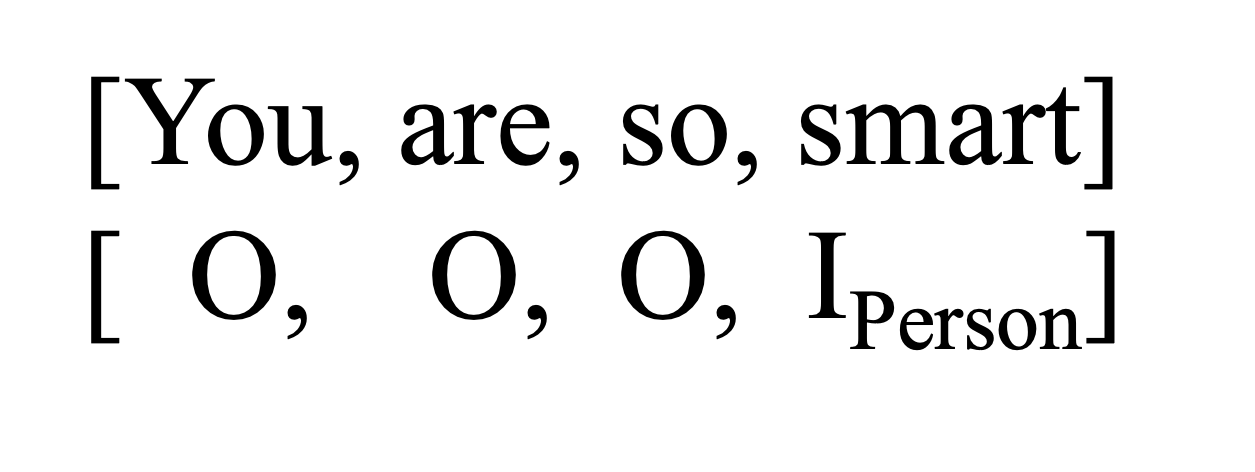}
\vspace{-0.5cm}
\caption{Labeling the praise components using IO scheme.} \label{fig:token_list}
\end{figure}

In line with prior study \cite{lin2024can}, we adopted the Inside-Outside (IO) labeling scheme, as described by Konkol et al. \cite{konkol2015segment}, to better analyze effective praise in tutoring dialogues. The IO scheme is particularly suited to our needs because it captures essential information without the complexity of marking entity boundaries. This scheme, known for its simplicity and efficiency, uses \textbf{I} tags to indicate praise components and \textbf{O} tags for non-praise words. For instance, in the phrase \textit{``You are so smart''}, the words \textit{``smart''} are labeled as part of the person-based praise (\textbf{$ \mathrm{I_{Person}} $}), while the remaining words are tagged as \textbf{O}, as illustrated in Figure \ref{fig:token_list}. This method allows us to focus on the core aspects of praise effectively.

\subsection{Fine-tuning GPTs with Augmented Data}

To answer our two \textbf{RQ}s, we fine-tuned GPT models using augmented data. We then assessed their classification accuracy to evaluate the effectiveness of the data augmentation process.

\subsubsection{Fine-tuning GPTs}
We utilized GPT-3.5 to evaluate the effectiveness of data augmentation. This approach involved fine-tuning the GPT-3.5 model to recognize and understand patterns related to identifying praise components in tutor responses. To prepare the data for fine-tuning, we converted the tutor responses and their corresponding tags into JSON format. This structured format aligns with the input style required by the GPT model. The details of our fine-tuning scheme can be found in Appendix A.

\subsubsection{Data Augmentation with GPT-4o}
To address the issue of low-resource in fine-tuning GPT-3.5 for tutor response scenarios, we leveraged more advanced model GPT-4o to perform data augmentation and increase the number of labeled responses. Our initial attempt to use GPT-4o directly for data augmentation revealed that the model tended to generate easy and typical responses and labels, which limited the fine-tuned model's ability to learn from difficult cases. This issue arises because generalized LLMs often produce responses that lack the diversity and complexity needed for nuanced educational feedback tasks. As a result, these typical outputs do not adequately challenge the model to improve its performance on more complex or less common feedback scenarios. To overcome these limitations, we developed a structured data augmentation approach using  GPT-4o. Our approach focuses on generating synthetic datasets that retain authentic sentence structures and expressions, which ensures that the augmented data is both semantically coherent and contextually relevant, providing a richer and more varied training set. By doing this, we aim to introduce variability while preserving the grammaticality and coherence of the text. Such targeted data augmentation leads to improved model performance by providing the student LLMs (i.e., GPT-3.5) with more relevant and contextually rich examples, ultimately resulting in more accurate and reliable performance.

\begin{figure*}[h]
\centering
\includegraphics[width=0.65\textwidth]{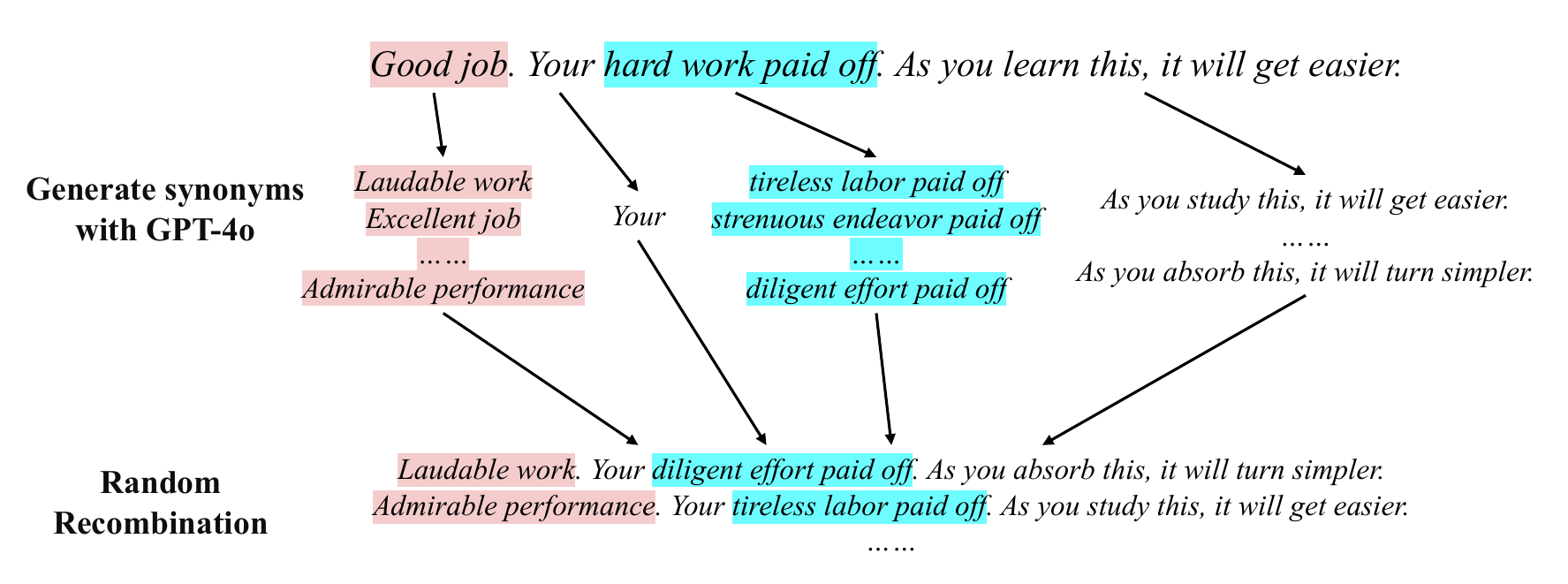}
\caption{Data augmentation process. Outcome-based praise (e.g., “Good job”) and effort-based praise (e.g., “Hard work paid off”) were diversified using GPT-4o to generate synonymous phrases, enabling the creation of varied responses to enhance model generalization.} \label{fig:augmentation}
\end{figure*}

To augment the data, we deconstructed the original responses to be augmented into their constituent components: effort-based praise, outcome-based praise, and person-based praise. As illustrated in Figure \ref{fig:augmentation}, outcome-based praise (\textit{``Good job''}, highlighted in \colorbox{outcome}{red}), and effort-based praise (\textit{``hard work paid off''}, highlighted in \colorbox{effort}{blue}). To increase the variety of expressions, we employed GPT-4o to generate synonymous phrases for each of these components. For example, \textit{``You did a perfect job!''} might be transformed into \textit{``Laudable work is done by you!''} or \textit{``Excellent job is achieved!''}, while \textit{``hard work paid off''} could become \textit{``tireless labor made sense''}. Subsequently, we reconstructed the responses by randomly recombining these synonymous phrases to form responses with similar meanings but different expressions. This method enabled us to create a more diverse training set, thereby improving the model's ability to generalize from limited data.

The prompt we used is shown in Table \ref{tab:prompt_augment}. The \textbf{TEXT} represents different components of the sentence to be augmented, including effort-based praise, outcome-based praise, person-based praise, and other non-praise components, as illustrated in Figure \ref{fig:augmentation}. To ensure the quality of the generated synonyms, we set the temperature parameter to 0. The temperature setting controls the variability of the model's output, and a value of 0 promotes more deterministic and high-quality responses \cite{openai_api}. 
During our prompt engineering process, we experimented with generating 5, 10, 15, 20, 25, and 30 synonyms. We found that generating 15 synonyms was optimal for ChatGPT-4o to produce high-quality synonyms without resorting to low-frequency words. For example, generating more than 15 synonyms for \textit{``You did a good job''} sometimes resulted in rare expressions like \textit{``You executed a commendable operation''}, which, while technically correct, is not commonly used in everyday educational settings. 
We tested five different prompts and observed consistent results across them. The quality of synthetic data is crucial, as demonstrated by our experiments with traditional data augmentation methods, which sometimes produced low-quality outputs (see Section 4.1.2 for results). In praise type labeling, even a single word difference can change the classification from effort-based to outcome-based praise. For instance, \textit{``You did a great job''} (outcome-based) versus \textit{``You did a great job persevering''} (effort-based) highlights how nuanced differences affect categorization. Therefore, high-quality data is essential for accurate model training.

\begin{table}[H]
\centering
\caption{Prompt for Data Augmentation}
\renewcommand{\arraystretch}{1.3}
\large
\resizebox{0.49\textwidth}{!}{%
\begin{tabular}{|lp{14cm}|}
\hline
\textbf{Role} & \textbf{Content} \\ \hline
\textbf{System} & \textit{You are required to rephrase the text in English through synonym replacement, ensuring the original context and meaning are preserved.} \\ 
\textbf{User} & \textit{Please note that the sentence structure and format must be preserved, with synonyms used only where they maintain the original meaning. \
        Retain words and ideas from the original response in English. Maintain similar lengths to the original text. \
        Please generate 15 unique sentences in English by applying synonym replacements to the text provided below.\
        One item per line, do not include numbers or bullet points. \
            Here is the text:} \textbf{TEXT}\\ 
\hline
\end{tabular}
\label{tab:prompt_augment}
}
\end{table}

\subsection{Metrics}
To assess the performance of our fine-tuned models, we employed three evaluation metrics: the $F_{2}$ score, the Intersection over Union score (IoU), and the modified Intersection over Union score (M-IoU). The formulas for these metrics are detailed below.

In sequence labeling tasks, traditional metrics like the $F_{1}$ score are commonly used to assess model performance \cite{brandsen2020creating}. However, the inclusion of additional entities did not hinder the trainees' ability to understand the correctness of the responses. For example, Table \ref{tab:different_highlight} illustrates that while expert annotations capture outcome-based praise accurately, model-generated annotations may include extra words (FP) that are still useful, unlike incomplete annotations (FN) that miss the praise's intent. This suggests a need for a metric that more flexibly accommodates additional praise tokens. Therefore, we propose modifying the Intersection over Union (IoU) concept, commonly used in computer vision, to better suit our evaluation needs.

\begin{table}[h]
\centering
\caption{Original and predicted praise with outcome-based praise in red and effort-based praise in blue.}
\renewcommand{\arraystretch}{1.5}
\label{tab:different_highlight}
\resizebox{0.29\textwidth}{!}{%
\begin{tabular}{clc}
\hline
    & \textbf{Instance}   & \textbf{Label}                                 \\ \hline
1         & \textit{Hey Kevin, you did a\colorbox{outcome}{great job}}. & True \\ \hline
2         & \textit{Hey Kevin, you did\colorbox{outcome}{a great job}}. & Pred \\ \hline
3         & \textit{Hey Kevin, you\colorbox{outcome}{did a great job}}. & Pred \\ \hline
4         & \textit{Hey Kevin,\colorbox{outcome}{you did a great job}}. & Pred \\ \hline
5         & \textit{Hey Kevin, you did a\colorbox{outcome}{great} job}. & Pred \\ \hline
\end{tabular}
}
\end{table}

The IoU metric (Equation \ref{eq:iou_score}), used in object detection and segmentation tasks, measures the overlap between predicted and actual annotations \cite{liu2021span, chen2023boundary}. In sequence labeling, the \textit{`Area of Overlap'} (TP) includes tokens correctly identified as praise, while the \textit{`Area of Union'} includes all tokens labeled as praise by the model (TP and FP) and all actual praise tokens in the groundtruth (TP and FN). The IoU metric inspired us in handling the challenge of additional detected words. To address additional detected words, we propose a metric that more flexibly accommodates additional praise tokens: Modified IoU (M-IoU) metric (Equation \ref{eq:miou_score}) that includes a weight coefficient, \(\alpha\), reducing the impact of FPs and adding flexibility. The \(\alpha\) value, set at 0.2 based on expert observations, allows adjustment of tolerance for additional words, which equates to a penalty level where five extra words are considered equivalent to one missing word. This reflects our approach's emphasis on being more tolerant of extra words while being less forgiving of missing words. For instance, if an output contains five unnecessary words, it incurs a penalty similar to having one crucial word missing, aligning with our dataset's annotation guidelines where annotators are instructed to mark the minimal words necessary to represent praise. As shown in Table \ref{tab:different_highlight}, including more extra words does not significantly affect the praise itself, whereas missing even one word can render it invalid. In cases where no praise is present and the model agrees (i.e., TP + FP + FN equals 0), we assign a score of 1 to reflect perfect agreement, showcasing the M-IoU's adaptability and effectiveness in practical applications. In our previous work \cite{lin2024can}, we validated the effectiveness of the M-IoU score for evaluating the quality of highlighted components from GPT models through correlation analysis. We also acknowledge that the $F_{2}$ score (Equation \ref{eq:f2_score}) can achieve a similar effect by assigning a weight of 0.2 to FP and 0.8 to FN. However, unlike M-IoU, the $F_{2}$ score does not provide the flexibility to adjust the balance between penalties for FP and FN tokens. For instance, with the $F_{2}$ score, approximately four extra words (FP) incur the same penalty as one missing word (FN), and this ratio is fixed. In contrast, M-IoU allows us to fine-tune this balance by adjusting the $\alpha$ coefficient. However, since the $F_{2}$ score is close to our M-IoU setup here ($\alpha = 0.2$), we have included it in our analysis as an additional metric to provide further insights.

\begin{equation}
\text{IoU} = \frac{\text{Area of Overlap}}{\text{Area of Union}} = \frac{TP}{TP + FP + FN} 
\label{eq:iou_score}
\end{equation}

\begin{equation}
\text{M-IoU} = \frac{TP}{TP + \alpha \times FP + FN}
\label{eq:miou_score}
\end{equation}

\begin{equation}
\begin{split}
\text{$F_{2}$ score} &= \frac{(1+2^2) \times TP}{(1+2^2) \times TP + FP + 2^2 \times FN} \\
                     &= \frac{TP}{TP + 0.2 \times FP + 0.8 \times FN}
\end{split}
\label{eq:f2_score}
\end{equation}

\section{Results}
\subsection{Results on RQ1}

To address \textbf{RQ1}, we aim to investigate the extent to which data augmentation methods can enhance the performance of fine-tuned large language models (LLMs) in providing explanatory feedback.

\subsubsection{Experimental Setup}
We begin by randomly extracting 13 labeled responses from the training set (consists of 65 labeled responses) to create a low-resource training set. This low-resource training set is then augmented using ChatGPT-4o under various data augmentation multipliers. These multipliers determine how many times the original number of responses is increased. For example, if the multiplier is 3 and we start with 5 responses, the augmented dataset will contain 15 responses. We evaluate the fine-tuned model's performance on the test set consisting of 64 labeled responses. The calculation of M-IoU, IoU, and F2 score was each repeated in 5 experiments, meaning a total of 15 experiments were repeated for each set size, with a different random seed used for each run.

\subsubsection{Analysis of Results}
The average M-IoU, IoU, $F_{2}$ score performance of the fine-tuned models, trained with the training set split with five different random seeds, is depicted in Table \ref{tab:effortres} and Table \ref{tab:outcomeres}.

\begin{table}[H]
\centering
\caption{Average Scores and Standard Errors for M-IoU, IoU, and $F_{2}$ Score (effort-based praise). 520* denotes traditional augmentation methods (e.g., synonym replacement, random word insertion) with a set size of 520.}
\resizebox{0.9\linewidth}{!}{
\begin{tabular}{ccS[table-format=1.3]S[table-format=1.3]S[table-format=1.3]S[table-format=1.3]S[table-format=1.3]S[table-format=1.3]S[table-format=1.3]S[table-format=1.3]}
\toprule
& & \multicolumn{2}{c}{\textbf{M-IoU}} & \multicolumn{2}{c}{\textbf{IoU}} & \multicolumn{2}{c}{\textbf{$F_{2}$ Score}}\\
\cmidrule(lr){3-4} \cmidrule(lr){5-6} \cmidrule(lr){7-8} \cmidrule(lr){9-10}
\textbf{Set Size} & & \textbf{Score} & \textbf{Error} & \textbf{Score} & \textbf{Error} & \textbf{Score} & \textbf{Error} \\
\midrule
13 & & 0.506 & 0.024 & 0.443 & 0.011 &0.541&0.027\\
26 & & 0.501 & 0.020 & 0.440 & 0.009 &0.547&0.034\\
65 & & 0.495 & 0.021 & 0.436 & 0.009 &0.540&0.025\\
130 & &0.551 & 0.018 & 0.488 & 0.013 &0.615&0.006\\
260 & &0.596 & 0.008 & 0.551 & 0.004 &0.636&0.012\\
520 & &0.601 & 0.006 & 0.554 & 0.007 &0.652&0.013\\
\midrule
520* & &0.456 & 0.036 & 0.392 & 0.014 &0.490&0.027\\
\bottomrule
\end{tabular}
}
\label{tab:effortres}
\end{table}

\begin{table}[H]
\centering
\caption{Average Scores and Standard Errors for M-IoU, IoU, and $F_{2}$ Score (outcome-based praise). 520* denotes traditional augmentation methods (e.g., synonym replacement, random word insertion) with a set size of 520.}
\resizebox{0.9\linewidth}{!}{
\begin{tabular}{ccS[table-format=1.3]S[table-format=1.3]S[table-format=1.3]S[table-format=1.3]S[table-format=1.3]S[table-format=1.3]S[table-format=1.3]S[table-format=1.3]}
\toprule
& & \multicolumn{2}{c}{\textbf{M-IoU}} & \multicolumn{2}{c}{\textbf{IoU}}  & \multicolumn{2}{c}{\textbf{$F_{2}$ Score}}\\
\cmidrule(lr){3-4} \cmidrule(lr){5-6} \cmidrule(lr){7-8} \cmidrule(lr){9-10}
\textbf{Set Size} & & \textbf{Score} & \textbf{Error} & \textbf{Score} & \textbf{Error} & \textbf{Score} & \textbf{Error} \\
\midrule
13 & & 0.623 & 0.022 & 0.597 & 0.023  &0.678& 0.015\\
26 & & 0.651 & 0.036 & 0.589 & 0.037  &0.710& 0.024\\
65 & & 0.672 & 0.014 & 0.642 & 0.020  &0.711& 0.018\\
130 & & 0.711 &0.017 & 0.669 & 0.021  &0.739& 0.006\\
260 & & 0.767 &0.001 & 0.752 & 0.001  &0.803& 0.003\\
520 & & 0.772 &0.002 & 0.755 & 0.001  &0.795& 0.003\\
\midrule
520* & &0.637 & 0.026 & 0.580 & 0.027 &0.684& 0.010\\
\bottomrule
\end{tabular}
}
\label{tab:outcomeres}
\end{table}

Table \ref{tab:effortres} presents the scores and errors for effort-based praise across different augmented training set sizes, ranging from 13 to 520 labeled responses, while Table \ref{tab:outcomeres} details the corresponding metrics for outcome-based praise. In both cases, as the training set size increases, there is a noticeable improvement in all three metrics—M-IoU, IoU, and $F_{2}$ scores. For effort-based praise, the M-IoU score starts at 0.506 with a set size of 13 and gradually increases to 0.601 with a set size of 520, with similar trends observed for the IoU and $F_{2}$ scores, indicating that larger augmented training sets contribute to better model performance. Similarly, for outcome-based praise, the M-IoU score begins at 0.623 for the smallest set size of 13 and reaches 0.772 with the largest set size of 520, with consistent improvements also seen in the IoU and $F_{2}$ scores. These results reinforce the effectiveness of data augmentation for enhancing the model's ability to accurately identify both effort-based and outcome-based praise. In addition to our primary methods, we employed traditional NLP data augmentation methods\textemdash random insertion, swap, deletion, and synonym replacement\textemdash to expand the original 13 responses into a dataset of 520 entries.  Specifically, each original response was augmented by applying these traditional methods collectively rather than generating separate datasets for each method individually. For example, a single tutor response would undergo synonym replacement, random word insertion, swapping, and deletion in combination, producing several augmented responses simultaneously. This combined augmentation approach allowed for more extensive dataset expansion by creating more diverse variants of each original response. The augmented dataset, containing a total of 520 responses, is labeled as '520*' in Table \ref{tab:effortres} and Table \ref{tab:outcomeres}, where we report the performance of models trained on this expanded dataset. However, these traditional data augmentation methods did not enhance performance and, in fact, resulted in worse outcomes compared to using no augmentation at all. This lack of improvement could be attributed to the introduction of incoherent or invalid sentences by these methods. For instance, the sentence \textit{``Hey Kevin, you did a great job.''} was altered to \textit{``Hey, Kevin, you the did an good work,''} which illustrates the potential for reduced clarity and grammatical correctness.

To statistically validate the observed improvements due to increasing the augmented dataset size, we performed the Mann-Whitney U test on key comparisons between data augmentation levels. For effort-based praise, although comparisons between the smallest sets (e.g., 13 vs. 26 and 13 vs. 65) did not show statistically significant differences (e.g., $U=13.00$, $p=0.5794$), larger increases in training data exhibited significant improvements; notably, comparing 13 vs. 260 responses yielded $U=4.00$ and $p=0.0476$, indicating significance at the 0.05 level. Similar trends were found when comparing intermediate augmentations such as 26 vs. 130 ($U=4.00$, $p=0.0476$) and 26 vs. 260 ($U=1.00$, $p=0.0079$), confirming that increasing data augmentation size significantly improves model performance.

For outcome-based praise, more pronounced statistical significance was observed. While smaller comparisons such as 13 vs. 26 ($U=9.00$, $p=0.2738$) did not reach significance, larger increments, including 13 vs. 130 ($U=2.00$, $p=0.0159$) and 13 vs. 260 ($U=0.00$, $p=0.0060$), demonstrated significant improvements. Additional comparisons such as 65 vs. 130 ($U=3.00$, $p=0.0278$) and 65 vs. 260 ($U=0.00$, $p=0.0060$) further confirmed these findings.

\begin{figure}[H]
\centering
\includegraphics[width=0.49\textwidth]{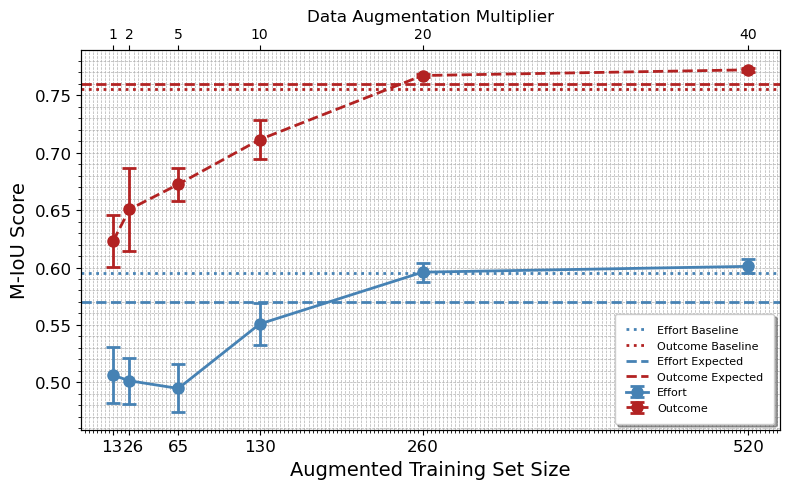}
\vspace{-0.4cm}
\caption{Performance of the fine-tuned GPT-3.5 model on highlighting correct types of praise with different augmented training set size.} \label{fig:result-rq1}
\end{figure}

As previously discussed, M-IoU offers a more flexible approach in accommodating additional praise tokens, which is particularly advantageous in our context. The inclusion of these additional entities did not impair the trainees' ability to comprehend the correctness of the responses. This flexibility is crucial for accurately capturing the model's performance across various scenarios. Furthermore, we selected M-IoU for this analysis to ensure consistency with our prior work \cite{lin2024can}, facilitating a more direct and meaningful comparison of the results.

Since we use expert annotations as groundtruth, we cannot directly calculate an M-IoU score for expert-level performance. However, based on our previous work \cite{lin2024can}, we estimate this score using a simple method. In our previous work, we validated the use of M-IoU as an evaluation metric. We engaged two trained human raters to assess the correctness of ChatGPT's annotations, scoring them from 1 (\textit{Strongly Disagree}) to 5 (\textit{Strongly Agree}). We then calculated the correlation between these human rating scores and the automatically calculated M-IoU scores, finding a strong positive correlation. This demonstrated that the M-IoU score aligns well with human judgment regarding annotation quality.
Given this strong correlation, we assumed a linear relationship between human scores and M-IoU. The same raters then scored expert annotations. We calculated the ratio between the human coders’ ratings on GPT predictions and their corresponding M-IoU scores, assuming this ratio remains consistent. By applying this ratio to the human coders’ ratings on expert annotations, we approximated the M-IoU score for expert performance, using it as a benchmark.
 This estimation provides a basic reference point: 0.57 for effort-based praise and 0.76 for outcome-based praise. If the model's performance exceeds these benchmarks, it suggests that the model performs at a level comparable to experts.

The average M-IoU performance of the fine-tuned models, trained with the training set split using five different random seeds, is depicted in Figure \ref{fig:result-rq1}. The x-axis of the graph represents the augmented training set size, ranging from 13 to 520, with corresponding data augmentation multipliers from 1 to 40. The augmented training set sizes—13, 26, 65, 130, 260, and 520—were chosen to follow a progressive doubling pattern, allowing us to systematically observe the impact of increasing data on model performance. The inclusion of 65 facilitates direct comparison with our previous work, while stopping at 520 was determined because the model achieved expert-level performance and showed minimal improvement compared to 260, making further increases unnecessary for this analysis. The y-axis shows the M-IoU score, reflecting the model's performance. The performance for effort-based praise is indicated by blue lines, while outcome-based praise is represented by red lines. The lines labeled \textbf{Effort} and \textbf{Outcome} depict the actual M-IoU scores obtained through the experiments. Error bars are included to show the variability across different random seeds, giving a sense of the robustness of the results. The horizontal dashed lines indicate the expected performance, aligning with expert annotation quality, while the horizontal dotted lines represent the baseline results, reflecting the best results from our previous work \cite{lin2024can} using the model fine-tuned with 65 labeled responses.

The results demonstrate that data augmentation positively impacts the model's performance for both outcome-based and effort-based praise, though with some differences in patterns. For outcome-based praise, the M-IoU score consistently increases from 0.623 to 0.772 as the augmented training set size grows from 13 to 520, surpassing both the best performance from our previous work and the expert annotation level. In contrast, effort-based praise shows a more complex trend: although there is an initial dip in the M-IoU score as the augmented set size increases from 13 to 65, the overall trend is positive, with the score eventually rising from 0.506 to 0.601 as the training set size reaches 520. This final score also exceeds the previous best results and expert-level annotations. In summary, despite some initial variability in effort-based praise, the overall results indicate that larger augmented training sets significantly enhance the model's ability to accurately highlight correct types of praise.

\begin{figure}[H]
\centering
\includegraphics[width=0.49\textwidth]{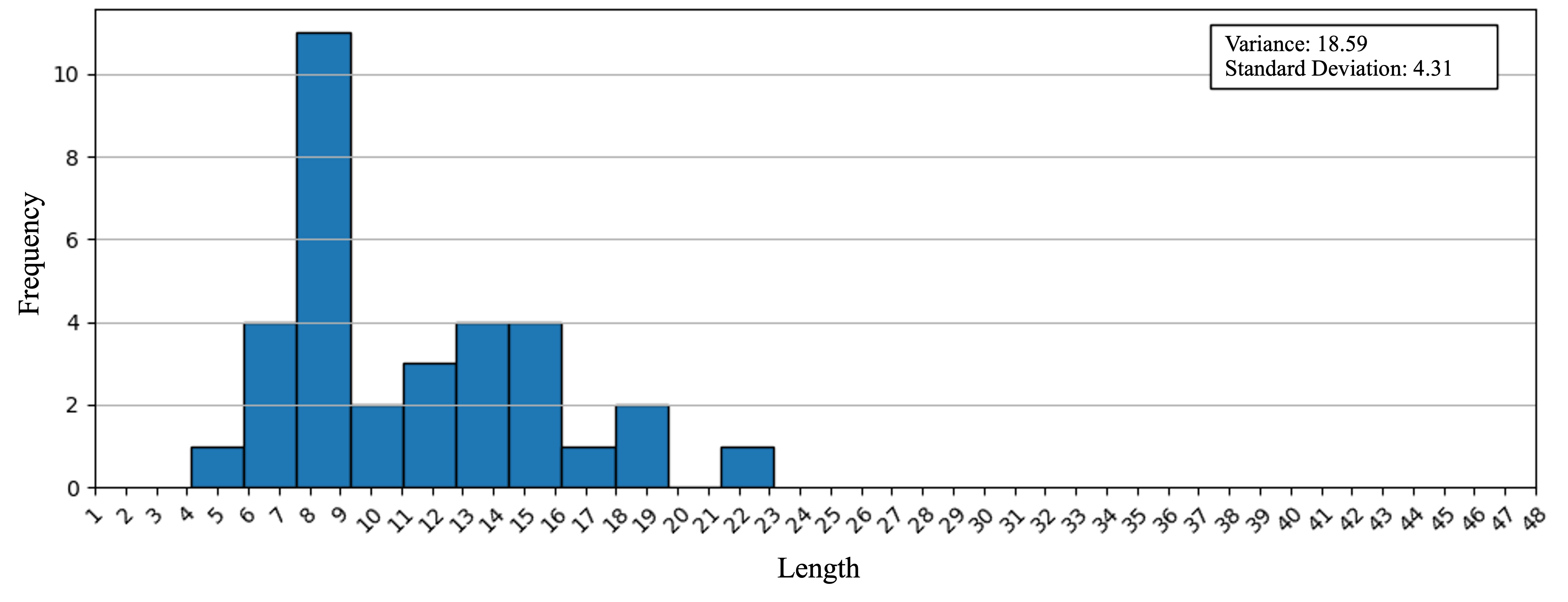}
\includegraphics[width=0.49\textwidth]{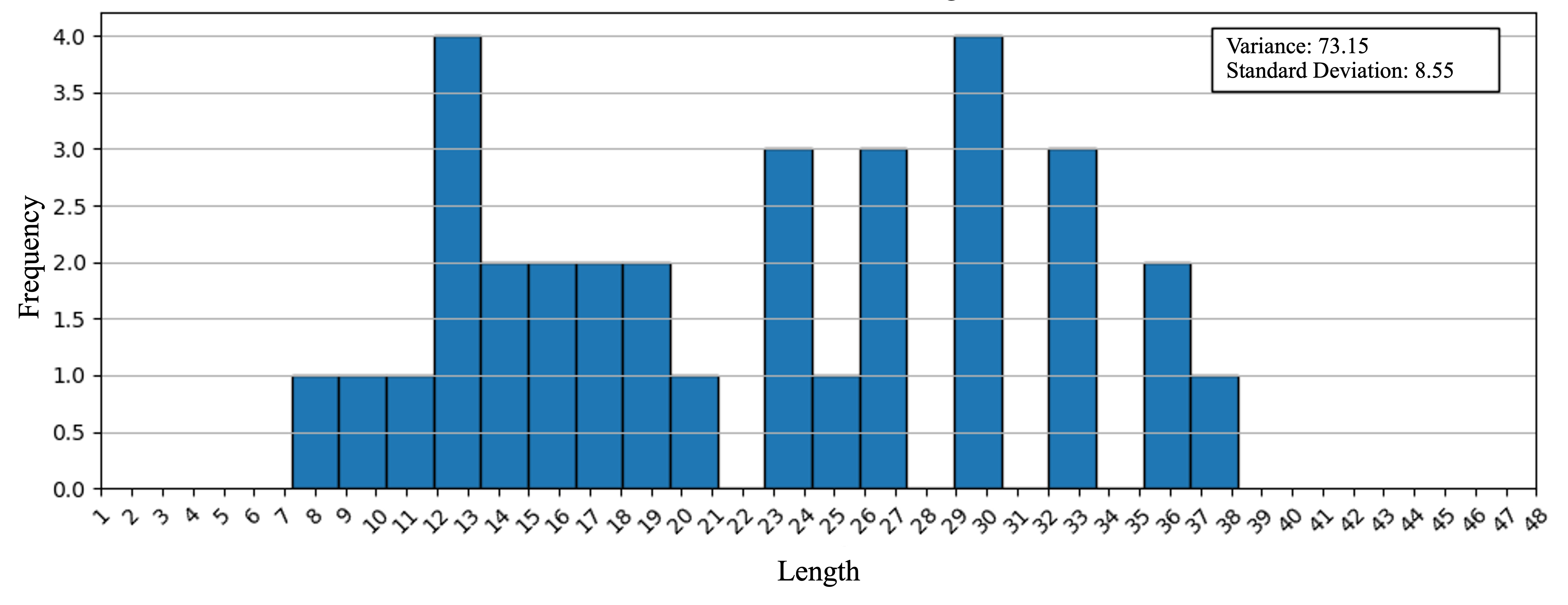}
\vspace{-0.2cm}
\caption{Word length distributions for outcome-based praise (top) and effort-based praise (bottom).} \label{fig:distribution}
\end{figure}

We observed from Figure \ref{fig:result-rq1} that, under the same level of data augmentation, the fine-tuned model performs worse on effort-based praise compared to outcome-based praise, despite the fact that outcome-based praise (with 33 distinct examples across all responses) is far less than effort-based praise (with 117 distinct examples across all responses), prompting us to investigate the reasons behind this difference. We randomly selected approximately 30 examples of both outcome-based and effort-based praise and analyzed the length statistics of outcome-based and effort-based praise. As depicted in Figure \ref{fig:distribution}, the length distribution of outcome-based praise is concentrated within a narrower and generally shorter range, while effort-based praise displays a wider range and greater variance in length.
The great variability in lengths of effort-based praise adds to the difficulty in accurately annotating entities, further complicating the model's performance. In contrast, the relatively lower diversity and shorter lengths in outcome-based praise allow the augmented data to effectively cover most information in the test set, resulting in higher performance.

\begin{figure}[h]
\centering
\includegraphics[width=0.39\textwidth]{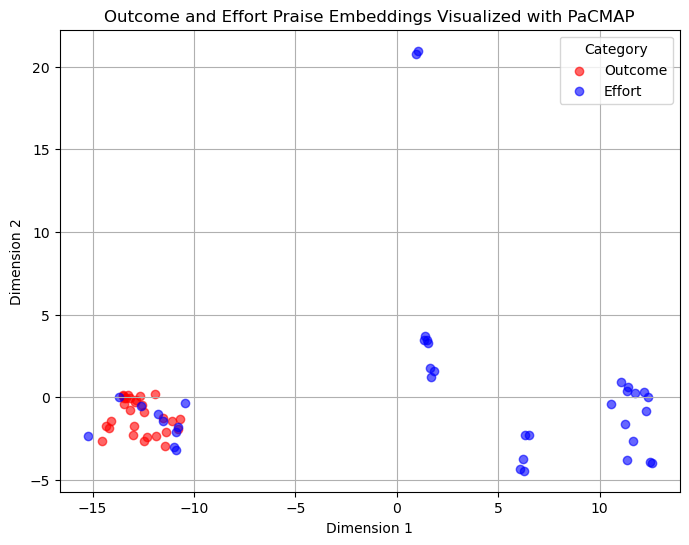}
\vspace{-0.2cm}
\caption{PaCMAP visualization of embedding spaces for outcome-based and effort-based praise.} \label{fig:tsne}
\end{figure}

We further investigated Semantic Textual Similarity \cite{majumder2016semantic} within effort-based and outcome-based praise, we used GTE models \cite{li2023towards} as our embedding model to convert all the praise into embeddings, and applied PaCMAP \cite{wang2021understanding} for 2-dimensional visualization to examine the distribution of these two types of praise. Figure \ref{fig:tsne} only shows the results under one random seed\footnote{For more results on the visualizations of outcome-based
and effort-based praise embeddings when praises are selected under different random seeds, please refer to Appendix E. }, which visualizes the embeddings of the two types of praise, revealing that effort-based praise spans a larger and more dispersed area in the semantic space than outcome-based praise. This greater semantic diversity within effort-based praise means that the augmented data only covers a limited portion of this extensive semantic space when the initial responses number is small, which causes the worse performance. While the narrow semantic space occupied by outcome-based praise allows the augmented data to cover most of the semantic space, leading to a better performance.

\subsection{Results on RQ2}

For \textbf{RQ2}, we aim to explore the extent to which our proposed data augmentation method can generalize to other types of praise, specifically, \textit{person-based praise}. 

\subsubsection{Experimental Setup and Data Imbalance}
As discussed in Section \ref{dataset}, we have a limited dataset with only 10 labeled responses featuring person-based praise. We combine these with our existing 129 labeled responses that do not include person-based praise, which results in a total of 139 labeled responses. Due to the significant data imbalance, where the majority of the praise is effort-based or outcome-based and only a small portion is person-based, we approach this as a two-class NER problem—instead of distinguishing between effort-based praise and outcome-based praise as in RQ1, we only require the model to identify person-based praise (labeled as \textbf{P}) and other parts (labeled as \textbf{O}). If we were to fine-tune the model on all types of praise without addressing this imbalance, the model might be biased towards predicting effort-based or outcome-based praise, given their predominance in the dataset. Following the method used in \textbf{RQ1}, we split the dataset into a training set and a test set. The training set consists of 70 labeled responses, including 5 person-based praise responses, while the test set contains 69 labeled responses, also including 5 person-based praise responses. We then apply our data augmentation method to both the training set and the test set. The augmentation multipliers were set to $\times 1, \times 2, \times 3$, and $\times 5$. The augmentation multipliers were set to $\times 1, \times 2, \times 3$, and $\times 5$ based on the model performance. In addition, we also augmented the person-based responses in the test set by $\times 20$ to obtain more precise evaluation values.

\subsubsection{Generalization Performance}
The results of this experiment are presented in Table \ref{table:result-rq2}.

% \begin{table}[H]
% \centering
% \caption{Average Scores and Standard Errors for M-IoU, IoU, and $F_{2}$ Score (person-based praise). \textbf{Aug. Size} represents \textbf{Augmented Training Set Size}, and \textbf{Mult.} represents \textbf{Augmentation Multiplier}.}
% \setlength{\tabcolsep}{3pt} % Adjust column spacing
% \renewcommand{\arraystretch}{1.1} % Adjust row spacing
% \begin{tabular}{ccS[table-format=1.3]S[table-format=1.3]S[table-format=1.3]S[table-format=1.3]S[table-format=1.3]S[table-format=1.3]}
% \toprule
% \multicolumn{2}{c}{\textbf{Aug. Size}} &  \multicolumn{2}{c}{\textbf{M-IoU}} & \multicolumn{2}{c}{\textbf{IoU}} & \multicolumn{2}{c}{\textbf{$F_{2}$ Score}}\\
% \cmidrule(lr){1-2}  \cmidrule(lr){3-4} \cmidrule(lr){5-6} \cmidrule(lr){7-8} \cmidrule(lr){9-10}
% \textbf{Mult.} & \textbf{Size} & \textbf{Score} & \textbf{Error} & \textbf{Score} & \textbf{Error} & \textbf{Score} & \textbf{Error} \\
% \midrule
% \times 1& 70& 0.808 & 0.014 & 0.797 & 0.016  &0.810&0.014\\
% \times 2& 140& 0.897 & 0.013 & 0.895 & 0.014 &0.902&0.014\\
% \times 3& 210& 0.966 & 0.017 & 0.965 & 0.013 &0.974&0.018\\
% \times 5& 350& 0.969 & 0.016 & 0.969 & 0.017 &0.976&0.016\\
% \bottomrule
% \end{tabular}
% \label{table:result-rq2}
% \end{table}

\begin{table}[H]
\centering
\caption{Average Scores and Standard Errors for M-IoU, IoU, and $F_{2}$ Score (person-based praise). \textbf{Aug. Size} represents \textbf{Augmented Training Set Size}, and \textbf{Mult.} represents \textbf{Augmentation Multiplier}.}
\setlength{\tabcolsep}{3pt} % Adjust column spacing
\renewcommand{\arraystretch}{1.1} % Adjust row spacing
\begin{tabular}{ccS[table-format=1.3]S[table-format=1.3]S[table-format=1.3]S[table-format=1.3]S[table-format=1.3]S[table-format=1.3]}
\toprule
\multicolumn{2}{c}{\textbf{Aug. Size}} &  \multicolumn{2}{c}{\textbf{M-IoU}} & \multicolumn{2}{c}{\textbf{IoU}} & \multicolumn{2}{c}{\textbf{$F_{2}$ Score}}\\
\cmidrule(lr){1-2}  \cmidrule(lr){3-4} \cmidrule(lr){5-6} \cmidrule(lr){7-8}
\textbf{Mult.} & \textbf{Size} & \textbf{Score} & \textbf{Error} & \textbf{Score} & \textbf{Error} & \textbf{Score} & \textbf{Error} \\
\midrule
$\times 1$ & 70  & 0.808 & 0.014 & 0.797 & 0.016 & 0.837 & 0.009 \\
$\times 2$ & 140 & 0.897 & 0.013 & 0.895 & 0.014 & 0.924 & 0.010 \\
$\times 3$ & 210 & 0.966 & 0.017 & 0.965 & 0.013 & 0.975 & 0.006 \\
$\times 5$ & 350 & 0.969 & 0.016 & 0.969 & 0.017 & 0.975 & 0.014 \\
\bottomrule
\end{tabular}
\label{table:result-rq2}
\end{table}

As is shown in Table \ref{table:result-rq2}, as the augmented training dataset size increases from 70 to 350, there is a clear upward trend across all three performance metrics: M-IoU, IoU, and $F_{2}$ Score. Specifically, the M-IoU score increases from 0.808 at a size of 70 to 0.969 at a size of 350. Similarly, the IoU score improves from 0.797 to 0.969, and the $F_{2}$ Score rises from 0.837 to 0.975. These results indicate that as we augment the training dataset, the model's ability to identify person-based praise improves consistently across all measured metrics. This gradual improvement demonstrates the effectiveness of our data augmentation method in enhancing the model’s performance.

Specifically, when the data augmentation multiplier is set to $\times 3$ (namely when the augmented training dataset size is 210), the model achieves satisfactory performance, the M-IoU, IoU, and $F_{2}$ scores already reached 0.965 or higher, and even achieved a perfect score of 1 under certain random seeds, suggesting that the model was nearing its performance ceiling, which demonstrates the model's robust ability to generalize to person-based praise. Beyond this multiplier, at $\times 5$, the performance plateaus, where increasing the multiplier to $\times 5$ yielded only marginal improvements, suggesting that further augmentation does not yield additional benefits.

The relatively high performance achieved with smaller augmentation multipliers (e.g., $\times 3$, namely the augmented training dataset size is 210) can be attributed to the nature of person-based praise, which often consists of straightforward adjectives describing a person's characteristics (e.g., \textit{``You are smart.''}, \textit{``You are the smartest student.''}). The simplicity in the vocabulary and word types—predominantly adjectives—makes it easier for the model to recognize and tag person-based praise, in contrast to the more nuanced effort-based and outcome-based praise categories examined in \textbf{RQ1}. This straightforward linguistic structure allows the model to detect patterns with greater ease, as opposed to the complexity found in the other categories, where varied and context-dependent language may pose more challenges for accurate tagging.

\section{Ablation Study}
In this section, we conduct three ablation studies to verify the effectiveness of our settings with outcome-based praise and effort-based praise. Since the goal here is not to compare the performance between outcome-based praise and effort-based praise, we report the average metrics across both types of praise for better and clearer comparison. All experiments were repeated 5 times under different random seeds.

\subsection{Augmenting Capabilities of GPT-4o and GPT-3.5}
To assess whether more advanced LLMs, such as GPT-4o, provide better data augmentation performance compared to GPT-3.5, we first evaluate the performance of data augmentation using GPT-4o and GPT-3.5, keeping all other settings consistent with the previous sections. The results are presented in Table \ref{tab:ab1}.

As shown in Table \ref{tab:ab1}, augmenting the dataset with GPT-4o consistently outperforms augmentation with GPT-3.5 across all metrics (M-IoU and IoU Score). This improvement can be explained by GPT-4o's ability to generate more valid and contextually appropriate synonym replacements compared to GPT-3.5. Additionally, we observed that GPT-3.5 often fails to adhere strictly to the clean answer structure we set in the prompt, leading to inconsistencies in the output format and introducing some noise into the augmented data. This issue was significantly less prevalent in outputs generated by GPT-4o, which produced more reliable and coherent augmentations.
For example, when augmenting a sentence like \textit{``You did a good job,''} GPT-3.5 might generate expressions such as \textit{``Sure, here is the synonym: You executed a commendable operation!''} which may introduce the noise \textit{``Sure, here is the synonym:''}.

\begin{table}[H]
\centering
\caption{Average M-IoU and IoU Score of augmentation with GPT-4o and GPT-3.5. We keep the base model as GPT-3.5.}
\resizebox{0.75\linewidth}{!}{
\begin{tabular}{ccS[table-format=1.3]S[table-format=1.3]S[table-format=1.3]S[table-format=1.3]S[table-format=1.3]S[table-format=1.3]S[table-format=1.3]S[table-format=1.3]}
\toprule
& & \multicolumn{2}{c}{\textbf{M-IoU}} & \multicolumn{2}{c}{\textbf{IoU}} & \\
\cmidrule(lr){3-4} \cmidrule(lr){5-6} \cmidrule(lr){7-8} 
\textbf{Set Size} & & \textbf{GPT-4o} & \textbf{GPT-3.5} & \textbf{GPT-4o} & \textbf{GPT-3.5}\\
\midrule
13 & & 0.565 & 0.554 & 0.520 & 0.493\\
26 & & 0.576 & 0.569 & 0.515 & 0.508\\
65 & & 0.584 & 0.574 & 0.539 & 0.521\\
130 & & 0.631 &0.598 & 0.579 & 0.546\\
260 & & 0.682 &0.601 & 0.652 & 0.550\\
520 & & 0.687 &0.609 & 0.654 & 0.561\\
\bottomrule
\end{tabular}
}
\label{tab:ab1}
\end{table}

\subsection{Initial Training Dataset Size}
In this section, we investigate the impact of different initial training dataset sizes on model performance after data augmentation. We use GPT-4o for data augmentation while keeping the base model as GPT-3.5. The results are presented in Table \ref{tab:ab2}, which shows the average M-IoU for training sets with different initial sizes when augmented to size 130, 260, and 520.

\begin{table}[H]
\centering
\caption{Average M-IoU for Training Sets with Different Initial Sizes after Augmentation.}
\resizebox{0.55\linewidth}{!}{
\begin{tabular}{ccccc}
\toprule
\textbf{Training Set Size} &  \multicolumn{3}{c}{\textbf{after Augmentation}} \\
\cmidrule(lr){2-4} 
\textbf{Initial}  & \textbf{130} & \textbf{260} & \textbf{520} \\
\midrule
13 &  0.631 & 0.682 & 0.687  \\
26 &  0.682 & 0.695 & 0.692 \\
65 &  0.692 & 0.690 & 0.691  \\
\bottomrule
\end{tabular}
}
\label{tab:ab2}
\end{table}

As shown in Table \ref{tab:ab2}, increasing the initial training set size leads to improved performance after augmentation, but the gains diminish as the augmented dataset size grows larger. For instance, starting with an initial set size of 13, the M-IoU improves from 0.631 to 0.687 as the augmented set size increases from 130 to 520. However, when starting with an initial set size of 65, the performance gains become minimal, suggesting that when the initial set size is relatively large, further augmentation may not be necessary or only a little augmentation is required. Beyond a certain point, additional augmentation yields diminishing returns.
This observation indicates that while data augmentation is effective in low-resource scenarios (e.g., starting with only 13 or 26 labeled instances), its impact becomes less pronounced as more labeled data is available initially. Therefore, for small datasets, augmentation plays a crucial role in boosting model performance, but its utility decreases as the dataset grows larger.

\subsection{Base Model}
In this section, we compare the performance of GPT-3.5 and GPT-4o as base models, while keeping GPT-4o as the augmentation model. The results are shown in Table \ref{tab:ab3}.

\begin{table}[H]
\centering
\caption{Average M-IoU and IoU Score with GPT-4o and GPT-3.5 as base model. We keep augmentation model as GPT-4o. }
\resizebox{0.75\linewidth}{!}{
\begin{tabular}{ccS[table-format=1.3]S[table-format=1.3]S[table-format=1.3]S[table-format=1.3]S[table-format=1.3]S[table-format=1.3]S[table-format=1.3]S[table-format=1.3]}
\toprule
& & \multicolumn{2}{c}{\textbf{M-IoU}} & \multicolumn{2}{c}{\textbf{IoU}} \\
\cmidrule(lr){3-4} \cmidrule(lr){5-6} 
\textbf{Set Size} & & \textbf{GPT-3.5} & \textbf{GPT-4o} & \textbf{GPT-3.5} & \textbf{GPT-4o}  \\
\midrule
13 & & 0.565 & 0.581 & 0.520 & 0.537 \\
26 & & 0.576 & 0.595 & 0.515 & 0.553\\
65 & & 0.584 & 0.622 & 0.539 & 0.584 \\
130 & & 0.631 &0.684 & 0.579 & 0.639  \\
260 & & 0.682 &0.689 & 0.652 & 0.657 \\
520 & & 0.687 &0.688 & 0.654 & 0.659\\
\bottomrule
\end{tabular}
}
\label{tab:ab3}
\end{table}

As shown in Table \ref{tab:ab3}, GPT-4o initially outperforms GPT-3.5 across all metrics (M-IoU, IoU, and $F_{2}$ Score) when the dataset size is small (e.g., with a set size of 13, GPT-4o achieves an M-IoU of 0.581 compared to 0.565 for GPT-3.5). This suggests that GPT-4o has stronger generalization capabilities with limited data, allowing it to perform better in low-resource scenarios.
However, as the dataset size increases, the performance gap between GPT-4o and GPT-3.5 begins to narrow, particularly at larger set sizes (e.g., at a set size of 520, both models achieve similar M-IoU scores of 0.687 for GPT-3.5 and 0.688 for GPT-4o). This indicates that while GPT-4o excels with smaller datasets due to its ability to generalize well from fewer examples, GPT-3.5 is able to catch up as more data becomes available.

\section{Discussion}

\subsection{Contributions}
In our study, we implemented a data augmentation approach using ChatGPT-4o to fine-tune the GPT-3.5 model for identifying and highlighting key components of trainee tutors' responses in low-resource scenarios. This approach demonstrated a significant improvement in model performance, which in turn facilitates the provision of automated feedback in tutor training programs. Our results show that with only 13 annotated instances, a model fine-tuned on synthetically augmented data can perform similarly to one trained on 65 annotated instances. This finding is particularly significant as it indicates a substantial reduction in the manual labor required for annotation, making the training process more efficient and scalable. These promising results suggest that the data augmentation techniques employed in this study could be leveraged to enhance model performance in other tutoring practices, such as responding to student errors and assessing students' knowledge \cite{lin2024rephrase}. This suggests the potential to develop automated feedback systems for various tutor training lessons, improving program efficacy and the tutoring process overall.

% \subsection{Generalization of Data Augmentation}
Then, we tested the generalization ability of our method by applying our data augmentation and fine-tuning method to person-based praise. The results indicate that even with a small number of labeled data, our method effectively boosts the model's performance. The effectiveness can be attributed to the ability of our proposed method to generate high-quality synthetic data that closely mimics real-world text distributions, allowing the model to learn robust patterns despite limited labeled data. Unlike other augmentation methods such as synonym replacement, our approach uses GPT-4o to maintain semantic coherence, capturing nuances crucial for effective model training.

To conclude, our contributions are two fold:

\begin{itemize}
    \item \textbf{Enhanced Model Performance}: Our data augmentation approach significantly improved the model's ability to identify various types of praise, achieving comparable results with fewer labeled instances.
    \item \textbf{Generalization Capability}: The method successfully extended to person-based praise, which is the most undesirable praise type, demonstrating its applicability to diverse types of feedback.
\end{itemize}

\subsection{Implications}

% \subsection{Generalization Beyond Tutor Feedback}
\textbf{Generalization beyond tutor feedback.} Although our primary focus was on generating automated explanatory feedback by highlighting key components in tutor trainees' responses for tutor training, the data augmentation methods we developed have broader applications in various educational contexts. For instance, tasks such as providing feedback on student essays often suffer from a lack of large labeled datasets \cite{nguyen2024learning}, particularly in low-resource settings like middle school classrooms \cite{jeon2024developing} or adult learning environments \cite{kleinert2009data}. Our augmentation approach, which enhances model performance with minimal labeled data, could be effectively integrated into these tasks to generate diverse training examples. This integration would enable models to better understand nuanced language and provide accurate assessments or feedback, even with limited labeled examples. As a result, it could reduce reliance on costly manual labeling while improving the quality of automated feedback systems.

\noindent\textbf{Enhancing GraphRAG for educational data mining through LLM-based data augmentation.} In addition to improving educational response labeling tasks, our study also has implications for building more advanced systems like GraphRAG (Graph-based Retrieval-Augmented Generation) \cite{edge2024local}. GraphRAG represents a promising direction for educational data mining because the GraphRAG combines knowledge graphs with retrieval-augmented generation to enhance entity extraction and reasoning under low-resource conditions. Our approach to data augmentation can be applied to GraphRAG by improving its ability to extract and represent entities from limited data sources. Specifically, our method of generating high-quality synthetic data could help GraphRAG models better capture nuanced relationships between entities, even when annotated data is scarce.
By integrating our augmentation techniques into GraphRAG, we can improve its performance in educational contexts where understanding complex relationships between concepts is crucial.

% \subsection{Model Performance with Data Augmentation and Semantic Richness}

\section{Limitation and Future Works}
Despite the demonstrated effectiveness of our data augmentation method, there are several limitations that need to be addressed. One major drawback is that our augmentation method, which modifies different parts of a response, is still constrained by the original structure of the initial data. This constraint limits the variety and diversity of the augmented data, potentially leading to less robust model fine-tuning. To overcome this limitation, future work could explore more sophisticated augmentation techniques, such as generative models that can create entirely new responses with varied structures while maintaining semantic coherence \cite{gandhi2024better}. 

Moreover, the current method may not fully capture the nuanced variations in natural language that are crucial for understanding complex feedback. For instance, slight differences in wording, such as \textit{``You did a great job''} (outcome-based praise) versus \textit{``You did a great job to persist''} (effort-based praise) can be different types of praise, which significantly impact the understanding of the response. To address this issue, future work could focus on developing a more sophisticated filtering policy. This policy would involve systematically evaluating and selecting augmented data based on specific criteria, such as semantic coherence, contextual relevance, and alignment with the desired feedback tone. By incorporating such a filtering mechanism, we can remove low-quality or less effective augmented data, ensuring that only the most contextually appropriate and high-quality data is used in training, thereby enhancing the overall effectiveness of the data augmentation process.

Another aspect worth exploring is the potential for more efficient methods of fine-tuning models in low-resource scenarios. For example, LoRA (Low-Rank Adaptation of Large Language Models) is a well-known technique for fine-tuning large models using a relatively small amount of data, such as hundreds of examples, which could be particularly effective in low-resource settings \cite{hu2021lora}. Investigating how such methods compare with our current approach could provide valuable insights into optimizing model performance when data is scarce.

Additionally, an alternative approach that could be explored in future research is the use of knowledge distillation from large language models such as GPT-4o \cite{viswanathan2023prompt2model}. Instead of relying on data augmentation, directly distilling knowledge from a highly capable model like GPT-4o could simplify the process while potentially improving data diversity. This approach may be more efficient and effective, particularly in enhancing the model's ability to generate a wider range of responses.

Furthermore, our current study focused primarily on a limited set of response types, specifically person-based, effort-based, and outcome-based praise. While our findings demonstrate the potential of our data augmentation method in these contexts, it is important to acknowledge that fully validating the generalizability and robustness of this approach requires testing on a broader range of response types. Future research should also explore its application to various other forms of educational feedback and responses, such as reacting to student errors and assessing students’ knowledge. By investigating how the model performs in highlighting key components of these additional feedback types, we can better understand its applicability across diverse educational scenarios. This broader exploration will further enhance the practical utility of our method in tutor training programs and beyond, providing insights into its potential for supporting a wide array of educational tasks.

In summary, while our data augmentation method shows promise in addressing low-resource problems, future work should aim to enhance the variety of augmented data and expand the scope of tested response types to further improve the robustness and applicability of the model. Additionally, our study did not systematically compare our approach with other data augmentation techniques, such as random insertion, swap, deletion, etc. Addressing these limitations in future research will allow us to develop more effective strategies for training models in diverse educational contexts, ensuring that our methods are both comprehensive and competitive.

\section{Conclusion}
In conclusion, this study demonstrated that leveraging advanced GPT models for data augmentation significantly improves the performance of explanatory feedback systems in low-resource educational settings. Our approach effectively addressed the data scarcity issue by generating high-quality synthetic labeled responses, thereby enhancing the model's ability to generalize across various types of praise. While the method showed robust performance, we identified areas for future improvement, including more sophisticated augmentation techniques and broader testing on diverse feedback types. Overall, our findings highlight the potential of data augmentation to reduce the dependency on large labeled datasets, making automated feedback systems more scalable and practical in diverse educational contexts.
In addition to exploring more sophisticated augmentation techniques and broader testing across diverse feedback types, future work will also focus on identifying and mitigating potential risks and biases introduced by synthetic data. Given that generative models may inadvertently reproduce or amplify existing biases, it is critical to develop robust evaluation and fairness auditing protocols to ensure that augmented datasets support equitable and reliable automated feedback systems in educational settings.

\section*{Acknowledgments}
This research was funded by the Richard King Mellon Foundation (Grant \#10851) and the Learning Engineering Virtual Institute (\href{https://learning-engineering-virtual-institute.org/}{https://learning-engineering-virtual-institute.org/}). The opinions, findings, and conclusions expressed in this paper are those of the authors alone.
We used ChatGPT to help revise the manuscript for grammar and fluency improvements, ensuring the text is clear and natural.

\appendix

\section{Appendix} 
\subsection{Details of the Fine-tuning Scheme} 
The fine-tuning of ChatGPT models involves several key steps to adapt the pre-trained model to specific tasks or domains. In our approach, we followed the standard process of fine-tuning ChatGPT\footnote{\url{https://platform.openai.com/docs/guides/fine-tuning}}. Each data point (one tutor response and its corresponding label) is structured into JSON format (see Table \ref{tab:prompt_design_finetuning} and \ref{tab:prompt_design_finetuning2}). The JSON format begins by presenting a \textbf{Lesson Principle} along with two labeled examples. The \textbf{Lesson Principle} provides key guidelines on how to praise students effectively, such as sincerity, specificity, immediacy, and a focus on the learning process (see details in Appendix \ref{lesson_prin_appendix1} and \ref{lesson_prin_appendix2}). Following this, two labeled examples are included as references to guide the model in applying the principles appropriately. In the \textbf{Tutor Response} field, we enter each tutor’s response that needs evaluation. The \textbf{Praise Type and Content} field includes labels indicating the type of praise (e.g., effort-based, outcome-based or person-based) and its content. Together, these components offer context and serve as few-shot examples, enabling fine-tuning. This process was repeated for each training data point, resulting in $k$ JSON data blocks, where $k$ is the total number of training examples.  Once the $k$ JSON data blocks were created, they were compiled to form the training set, which was then submitted to the GPT fine-tuning API using the default hyperparameters and settings.

\begin{table}[H]
\centering
\caption{Input for fine-tuning GPT 3.5 model (\textbf{RQ1})}
\renewcommand{\arraystretch}{1.2}
\resizebox{0.49\textwidth}{!}{%
\begin{tabular}{|lp{14cm}|}
\hline
\textbf{Role} & \textbf{Content} \\ \hline
\textbf{System} & \textit{You are a response evaluator designed to output JSON. Your task is to analyze tutor responses based on the principles of effective praise focusing on `effort' and `outcome'. Extract words or phrases that represent praise for the student's effort and outcome, and output the results in JSON format with keys titled `effort' and `outcome'.} \\ 
\textbf{User} & \textbf{Lesson Principle for \textbf{RQ1}} \\ 
\textbf{Assistant} & \textit{Sure, can you provide a tutor response for analysis} \\ 
\textbf{User} & \textit{Great job! You are a genius!} \\ 
\textbf{Assistant} & \textit{\{``effort'': [], ``outcome'': [``Great job'']\}} \\ 
\textbf{User} & \textit{Nice, let's do it again.} \\
\textbf{Assistant} & \textit{Sure, can you provide a tutor response for analysis?} \\ 
\textbf{User} & \textit{You are almost there! I am proud of how you are persevering through and striving to solve the problem. Keep going!}\\ 
\textbf{Assistant} & \textit{\{``effort'': [``persevering through and striving to solve the problem'', ``Keep going''], ``outcome'': []\}} \\ 
\textbf{User} & \textit{Nice, let's do it again.} \\ 
\textbf{Assistant} & \textit{Sure, can you provide a tutor response for analysis} \\ 
\textbf{User} & \textbf{Tutor Response} \\
\textbf{Assistant} & \textbf{Praise Type and Content} \\ 
\hline
\end{tabular}
\label{tab:prompt_design_finetuning}
}
\end{table}

\begin{table}[H]
\centering
\caption{Input for fine-tuning GPT 3.5 model (\textbf{RQ2})}
\renewcommand{\arraystretch}{1.2}
\resizebox{0.49\textwidth}{!}{%
\begin{tabular}{|lp{14cm}|}
\hline
\textbf{Role} & \textbf{Content} \\ \hline
\textbf{System} & \textit{You are a response evaluator designed to output JSON. Your task is to analyze tutor responses based on the principles of effective praise focusing on undesired part 'person\_based praise'. Extract words or phrases that represent person-based praise for the student's, and output the results in JSON format with keys titled 'person'.} \\ 
\textbf{User} & \textbf{Lesson Principle for \textbf{RQ2}} \\ 
\textbf{Assistant} & \textit{Sure, can you provide a tutor response for analysis} \\ 
\textbf{User} & \textit{You are smart and capable of completing this assignment with success.} \\ 
\textbf{Assistant} & \textit{\{"person": ["smart"]\}} \\ 
\textbf{User} & \textit{Nice, let's do it again.} \\
\textbf{Assistant} & \textit{Sure, can you provide a tutor response for analysis?} \\ 
\textbf{User} & \textit{Fantastic job, Kevin! You are the smartest student I have ever met! I wish all students were as smart as you.}\\ 
\textbf{Assistant} & \textit{\{"person": ["smartest student", "as smart as you"]\}} \\ 
\textbf{User} & \textit{Nice, let's do it again.} \\ 
\textbf{Assistant} & \textit{Sure, can you provide a tutor response for analysis} \\ 
\textbf{User} & \textbf{Tutor Response} \\
\textbf{Assistant} & \textbf{Praise Type and Content} \\ 
\hline
\end{tabular}
\label{tab:prompt_design_finetuning2}
}
\end{table}

\subsection{Lesson Principles for \textbf{RQ1}} 
\label{lesson_prin_appendix1}
The following is the principle that a correct response should follow:
  
Praising students for working hard and putting forth effort is a great way to increase student motivation. When the learning gets tough, giving correct praise is a powerful strategy to encourage students to keep going.

The correct response should be: \\
\hspace*{0.5cm}- perceived as sincere, earned, and truthful.\\
\hspace*{0.5cm}- specific by giving details of what the student did well.\\
\hspace*{0.5cm}- immediate with praise given right after the student action.\\
\hspace*{0.5cm}- authentic and is not repeated often, such as “great job” which loses meaning and becomes predictable.\\
\hspace*{0.5cm}- focused on the learning process, not ability\\
Correct responses must follow some, but not all the above. \\
There are two types of praise responses: Effort and Outcome praise \\
\hspace*{0.5cm}- Effort praise focuses on the learning process. Effort praise recognizes students for putting forth effort and persevering through the learning process instead of focusing on whether a student got the problem correct or pure ability.\\
\hspace*{0.5cm}- Outcome praise showcases student's achievements, such as getting a grade A on an assignment or getting a problem correct, and is often, but not always, associated with unproductive praise.

To receive full credit of correct praise, tutors cannot just say ``great job'' and praise with no specific reasoning. Tutors need to praise for effort AND be positive and encouraging.

\subsection{Lesson Principles for \textbf{RQ2}} 
\label{lesson_prin_appendix2}
The following is the principle that a correct response should follow:
  
Praising students for working hard and putting forth effort is a great way to increase student motivation. When the learning gets tough, giving correct praise is a powerful strategy to encourage students to keep going.

The correct response should be : \\
\hspace*{0.5cm}- perceived as sincere, earned, and truthful.\\
\hspace*{0.5cm}- specific by giving details of what the student did well.\\
\hspace*{0.5cm}- immediate with praise given right after the student action.\\
\hspace*{0.5cm}- authentic and is not repeated often, such as “great job” which loses meaning and becomes predictable.\\
\hspace*{0.5cm}- focused on the learning process, not ability\\
Correct responses must follow some, but not all the above. \\
There is one type of praise responses should be avoided: Person-based praise \\
\hspace*{0.5cm}- Person-based praise attributes success to innate qualities beyond the student's control and is often considered less effective. 

To receive full credit of correct praise, tutors cannot just say "great job" and praise with no specific reasoning. Tutors need to praise for effort AND be positive and encouraging.

\subsection{Distributions of Outcome-based and Effort-based Praise Lengths}
\label{fig:moredistribution}
This section presents additional results on the distributions of outcome-based and effort-based praise lengths when praises are selected under different random seeds. Since there are only 33 unique outcome-based praises, we included all of them and randomly selected 33 effort-based praises for comparison.

\begin{figure}[H]
\centering
\includegraphics[width=0.45\textwidth]{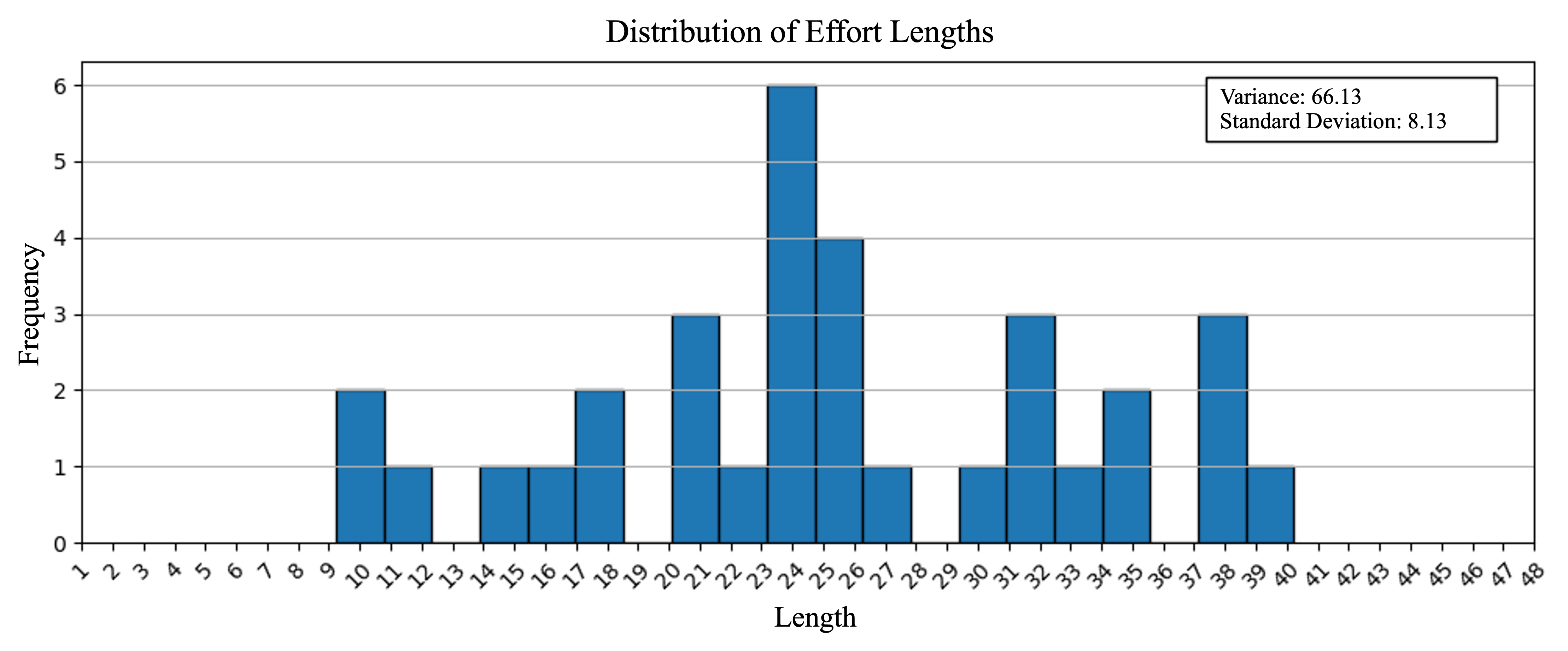}
\includegraphics[width=0.45\textwidth]{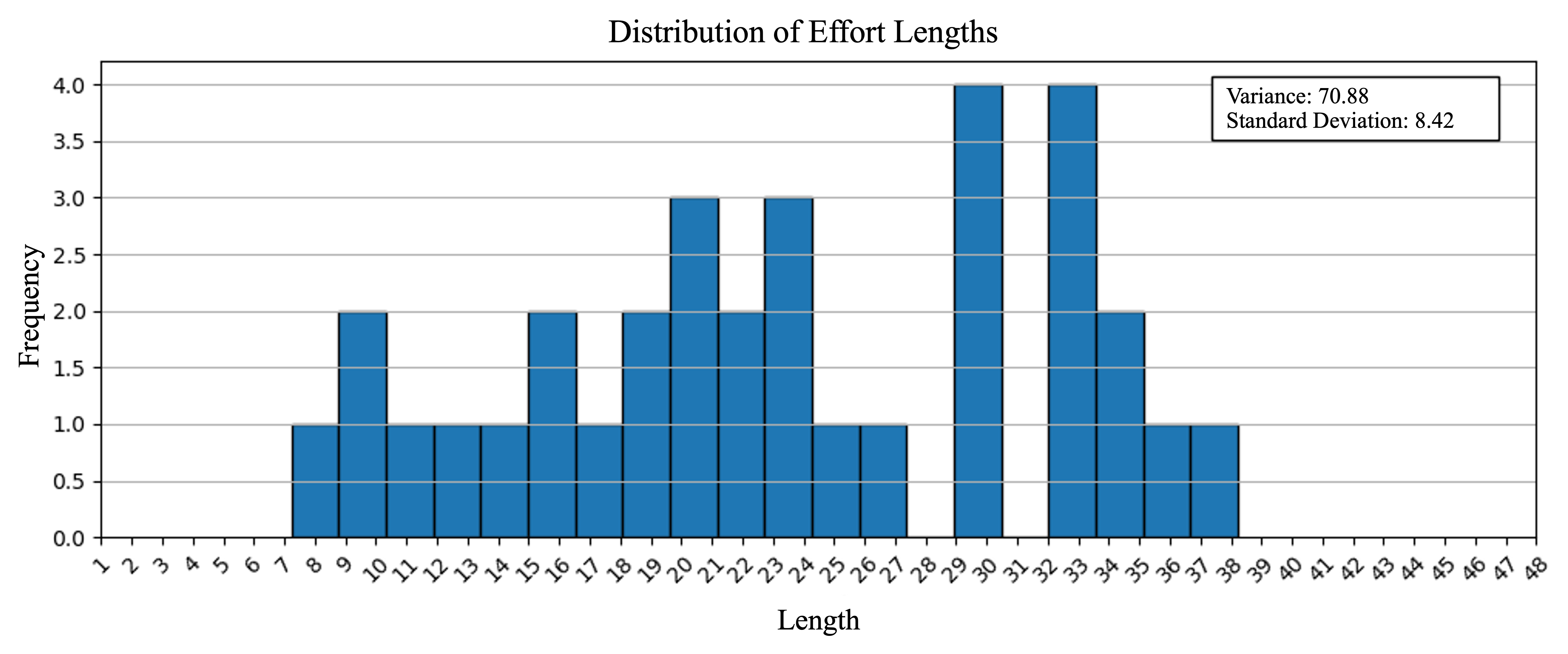}
\includegraphics[width=0.45\textwidth]{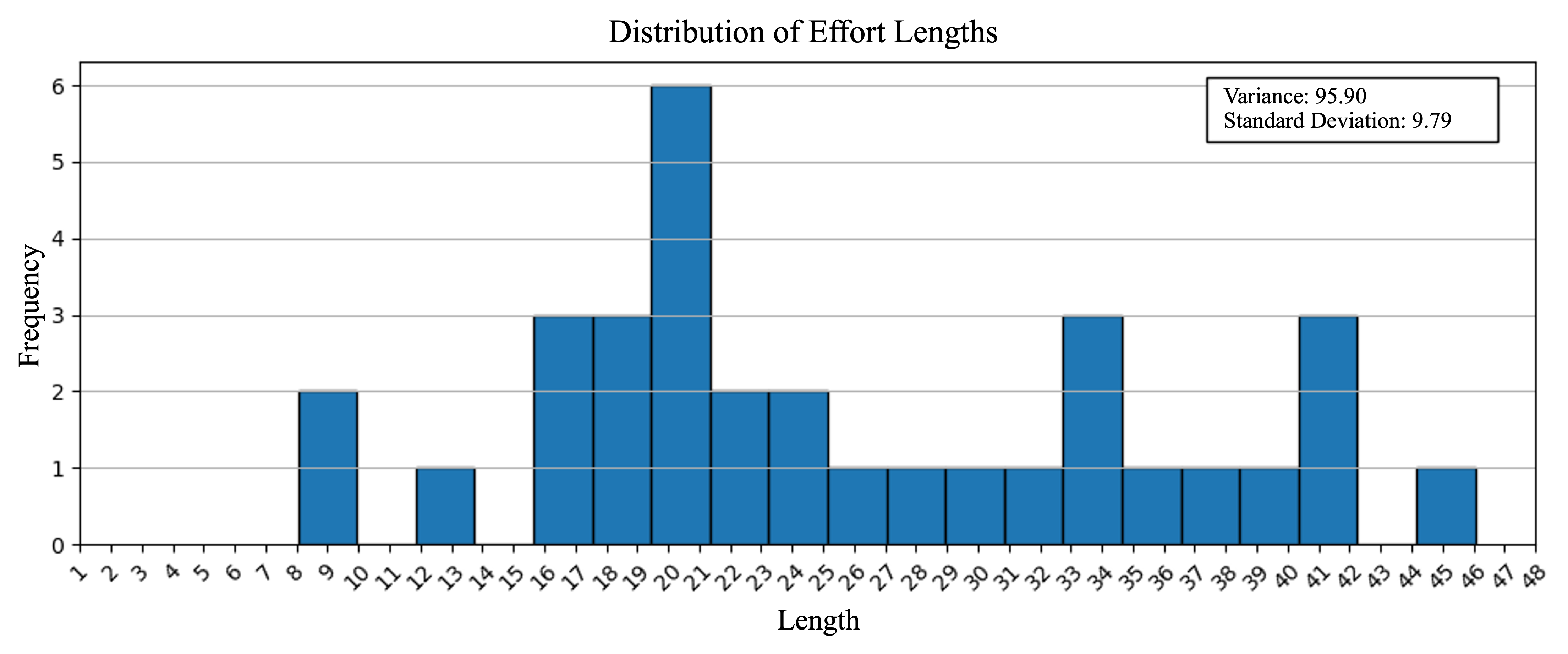}
\includegraphics[width=0.45\textwidth]{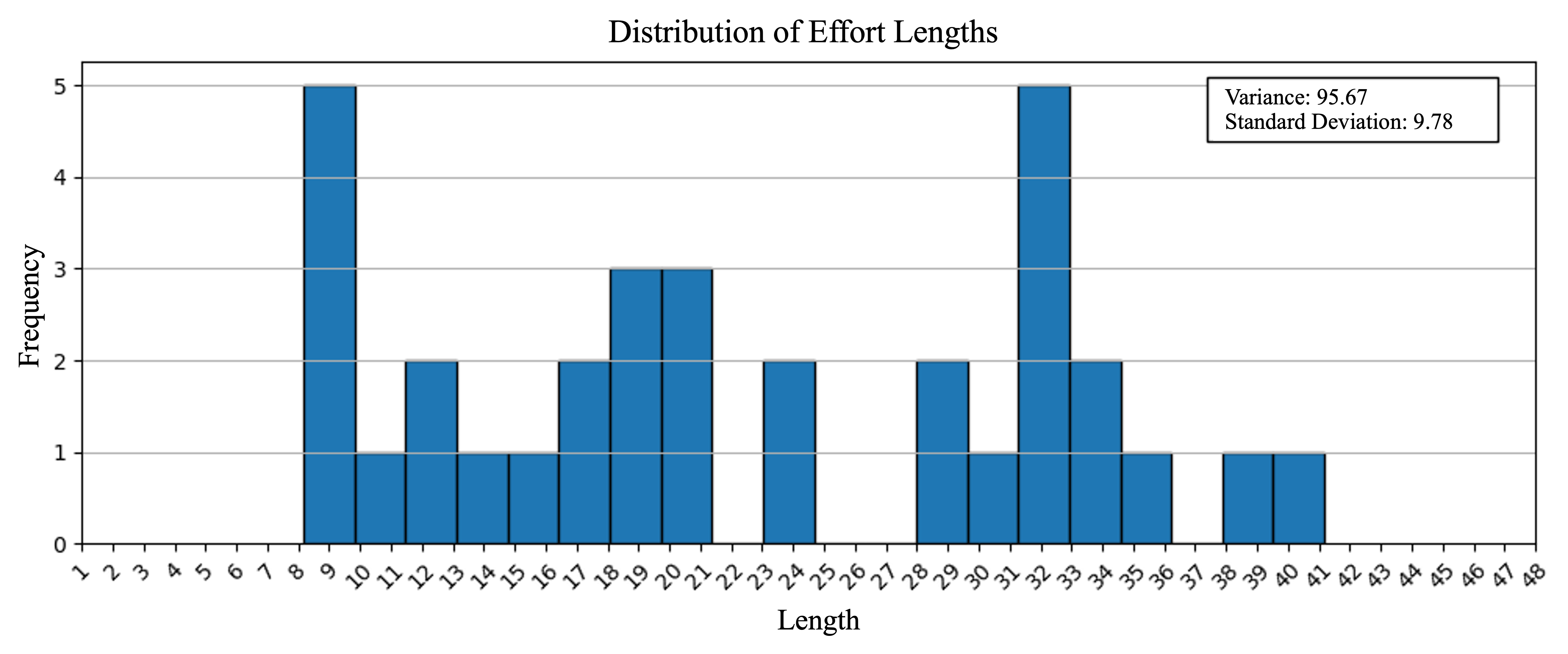}
\vspace{-0.2cm}
\caption{Distributions of Effort-based Praise Lengths (under more random seeds).} 
\end{figure}

\subsection{Outcome-based and Effort-based Praise Embeddings Visualized with PaCMAP} 
\label{fig:moretsne}
This section presents additional results on the visualizations of outcome-based and effort-based praise embeddings when praises are selected under different random seeds. Since there are only 33 unique outcome-based praises, we included all of them and randomly selected 33 effort-based praises for comparison.

\begin{figure}[H]
\centering
\includegraphics[width=0.38\textwidth]{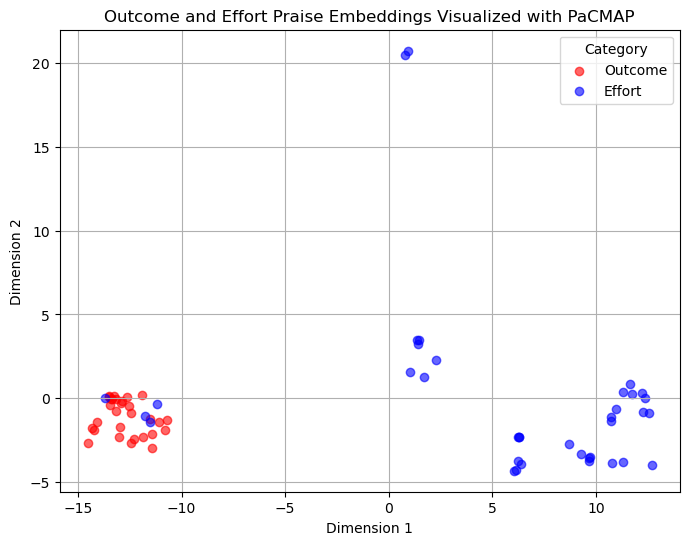}
\end{figure}

\begin{figure}[H]
\centering
\includegraphics[width=0.38\textwidth]{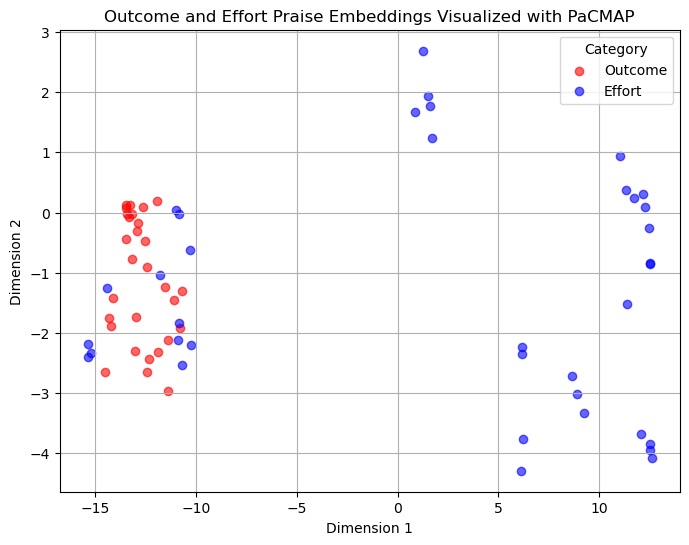}
\includegraphics[width=0.38\textwidth]{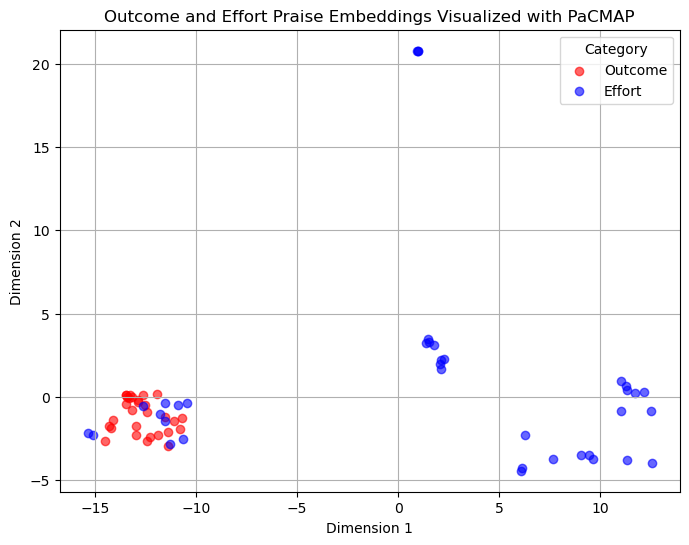}
\includegraphics[width=0.38\textwidth]{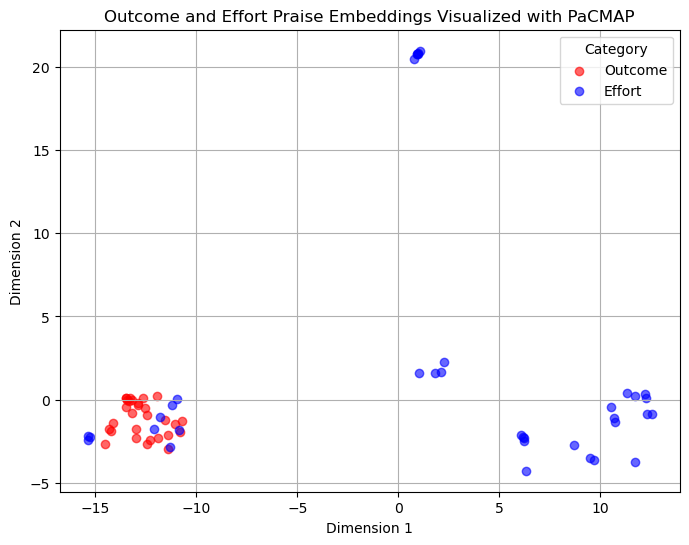}
\vspace{-0.2cm}
\caption{Outcome-based and Effort-based Praise Embeddings Visualized with PaCMAP (under more random seeds).} 
\end{figure}

% \begin{thebibliography}{99}
\bibliographystyle{IEEEtran}

\bibliography{main}

% \bibitem{ref1}
% {\it{Mathematics Into Type}}. American Mathematical Society. [Online]. Available: https://www.ams.org/arc/styleguide/mit-2.pdf

% \bibitem{ref2}
% T. W. Chaundy, P. R. Barrett and C. Batey, {\it{The Printing of Mathematics}}. London, U.K., Oxford Univ. Press, 1954.

% \bibitem{ref3}
% F. Mittelbach and M. Goossens, {\it{The \LaTeX Companion}}, 2nd ed. Boston, MA, USA: Pearson, 2004.

% \bibitem{ref4}
% G. Gr\"atzer, {\it{More Math Into LaTeX}}, New York, NY, USA: Springer, 2007.

% \bibitem{ref5}M. Letourneau and J. W. Sharp, {\it{AMS-StyleGuide-online.pdf,}} American Mathematical Society, Providence, RI, USA, [Online]. Available: http://www.ams.org/arc/styleguide/index.html

% \bibitem{ref6}
% H. Sira-Ramirez, ``On the sliding mode control of nonlinear systems,'' \textit{Syst. Control Lett.}, vol. 19, pp. 303--312, 1992.

% \bibitem{ref7}
% A. Levant, ``Exact differentiation of signals with unbounded higher derivatives,''  in \textit{Proc. 45th IEEE Conf. Decis.
% Control}, San Diego, CA, USA, 2006, pp. 5585--5590. DOI: 10.1109/CDC.2006.377165.

% \bibitem{ref8}
% M. Fliess, C. Join, and H. Sira-Ramirez, ``Non-linear estimation is easy,'' \textit{Int. J. Model., Ident. Control}, vol. 4, no. 1, pp. 12--27, 2008.

% \bibitem{ref9}
% R. Ortega, A. Astolfi, G. Bastin, and H. Rodriguez, ``Stabilization of food-chain systems using a port-controlled Hamiltonian description,'' in \textit{Proc. Amer. Control Conf.}, Chicago, IL, USA,
% 2000, pp. 2245--2249.

% \end{thebibliography}

\vfill

\end{document}